\documentclass[12pt, a4paper]{article}
\pdfoutput=1

\usepackage{amsmath,amssymb,amsfonts}
\usepackage{xfrac}
\usepackage{pifont}
\usepackage{physics}
\usepackage{bm}
\usepackage{bbm}
\usepackage{enumitem}
\usepackage{multirow}
\usepackage[utf8]{inputenc}
\usepackage{cite}
\usepackage[footnotesize]{caption}
\usepackage{slashed}

\usepackage{ifpdf}
\ifpdf        
  \usepackage{graphicx, hyperref, xcolor}     
\else     
  \usepackage[dvipdfmx]{graphicx, hyperref, xcolor}     
 \fi
\usepackage{subcaption}

\usepackage{tikz}
\usepackage{tikz-feynman}
\tikzfeynmanset{compat=1.0.0}
 
\definecolor{rossoferrari}{HTML}{D9073D}
\definecolor{mediumblue}{HTML}{0000CD}
\definecolor{forestgreen}{HTML}{228B22}
\definecolor{desy_blue}{HTML}{009EE2}
\definecolor{desy_orange}{HTML}{FD8800}
\definecolor{light_blue}{rgb}{0.284602,0.317763,0.963947}
\hypersetup{
setpagesize=false,
bookmarksnumbered=true,%
bookmarksopen=true,%
colorlinks=true,%
linkcolor=rossoferrari,
urlcolor=mediumblue,
citecolor=mediumblue,
linktocpage=false,
}

\usepackage[height=8.85in,width=6.8in]{geometry}

\usepackage{fourier}

\makeatletter
\@addtoreset{equation}{section}

\makeatother

\begin{document}

\begin{titlepage}

\begin{flushright}
TU-1140\\
KEK-TH-2379 \\
UMN-TH-4112/22 \\
FTPI-MINN-22-01
\end{flushright}

\begin{center}

\hfill 

\vskip .5in

{\LARGE \bf
Scalar field couplings to quadratic curvature \vspace{2.5mm}\\ 
and decay into gravitons\\
}

\vskip .8in

{\large Yohei Ema$^{a}$, Kyohei Mukaida$^{b,c}$, Kazunori Nakayama$^{d,e,f}$}

\vskip .3in
{\begin{tabular}{ll}
$^a$& \!\!\!\!\!\emph{William I. Fine Theoretical Physics Institute, School of Physics and Astronomy,}\\[-.15em]
& \!\!\!\!\!\emph{University of Minnesota, Minneapolis, MN 55455, USA}\\
$^b$& \!\!\!\!\!\emph{Theory Center, IPNS, KEK, 1-1 Oho, Tsukuba, Ibaraki 305-0801, Japan}\\
$^c$& \!\!\!\!\!\emph{Graduate University for Advanced Studies (Sokendai),}\\
$^{ }$& \!\!\!\!\!\emph{1-1 Oho, Tsukuba, Ibaraki 305-0801, Japan}\\
$^d$& \!\!\!\!\!\emph{Department of Physics, Tohoku University, Sendai, Miyagi 980-8578, Japan}\\
$^e$& \!\!\!\!\!\emph{International Center for Quantum-field Measurement Systems for Studies of}\\
$^{ }$& \!\!\!\!\!\emph{the Universe and Particles (QUP), KEK, 1-1 Oho, Tsukuba, Ibaraki 305-0801, Japan}\\
$^f$& \!\!\!\!\!\emph{Kavli IPMU (WPI), University of Tokyo, Kashiwa 277-8583, Japan}\\
\end{tabular}}

\end{center}
\vskip .6in

\begin{abstract}
\noindent
Any field, even if it lives in a completely hidden sector, interacts with the visible sector at least via the gravitational interaction.
In this paper, we show that a scalar field in such a hidden sector generically couples to the quadratic curvature via dimension five operators, \textit{i.e.}, $\phi R^2$, $\phi R^{\mu\nu} R_{\mu\nu}$, $\phi R_{\mu\nu\rho\sigma} R^{\mu\nu\rho\sigma}$, $\phi R_{\mu\nu\rho\sigma} \tilde R^{\mu\nu\rho\sigma}$,
because they can emerge if there exist particles heavier than it.
We derive these scalar couplings to the quadratic curvature by integrating out heavy particles in a systematic manner.
Such couplings are of phenomenological interest since some of them induce the scalar decay into the graviton pair.
We point out that the decay of a scalar field can produce a substantial amount of the stochastic cosmic graviton background at high frequency since the suppression scale of these operators is given by the mass of heavy particles not by the Planck scale.
The resultant graviton spectrum is computed in some concrete models.

\end{abstract}

\end{titlepage}

\renewcommand{\thepage}{\arabic{page}}
\setcounter{page}{1}

\tableofcontents
\renewcommand{\thepage}{\arabic{page}}
\renewcommand{\thefootnote}{$\natural$\arabic{footnote}}
\setcounter{footnote}{0}

\newpage
\section{Introduction}

Among the four fundamental forces, the gravitational interaction is special for its universality.
Any ingredients, including the standard model (SM) and possible dark sectors solving the puzzles in the SM, should at least interact through gravity.
Recently, there is a revival of attention to a dark sector that interacts with the SM only through gravity~\cite{Ema:2015dka,Garny:2015sjg,Markkanen:2015xuw,Ema:2016hlw,Schiappacasse:2016nei,Tang:2016vch,Tang:2017hvq,Garny:2017kha,Ema:2018ucl,Garny:2018grs,Chung:2018ayg,Hashiba:2018tbu,Ema:2019yrd,Li:2019ves,Cembranos:2019qlm,Herring:2020cah,Ahmed:2020fhc,Ema:2020ggo,Karam:2020rpa,Kolb:2020fwh,Gross:2020zam,Ling:2021zlj,Mambrini:2021zpp,Basso:2021whd}.
Most of the studies focus on the gravitational interaction within the Einstein gravity or the nonminimal coupling to gravity, namely up to the dimension four operators.
However, if the dark sector involves a scalar field, it can couple to the quadratic curvature through dimension five operators, such as $\phi R^2$, $\phi R^{\mu\nu} R_{\mu\nu}$, $\phi R_{\mu\nu\rho\sigma} R^{\mu\nu\rho\sigma}$, and $\phi R_{\mu\nu\rho\sigma} \tilde R^{\mu\nu\rho\sigma}$.
These operators lead to unique phenomena absent in the standard gravity~\cite{Antoniadis:1993jc,Kanti:1998jd,Nojiri:2005vv,Lue:1998mq,Choi:1999zy,Alexander:2004us,
Alexander:2004wk,Lyth:2005jf,Fischler:2007tj,DeSimone:2016bok,Kawai:2017kqt,Kamada:2020jaf}.
In particular, the last two operators, 
\begin{align}
	&\mathcal L_{\phi RR} = -\frac{\phi}{\Lambda}R_{\mu\nu\rho\sigma}R^{\mu\nu\rho\sigma},\label{phiRR} \\
	&\mathcal L_{\phi R\widetilde R} = -\frac{\phi}{\Lambda}R_{\mu\nu\rho\sigma}\widetilde R^{\mu\nu\rho\sigma}, \label{phiRtilR}
\end{align}
induce a decay of $\phi$ into graviton pairs (see, \emph{e.g.},~\cite{Delbourgo:2000nq,Alonzo-Artiles:2021mym}), 
which is never realized in the Einstein gravity nor the nonminimal coupling to gravity~\cite{Ema:2015dka,Delbourgo:2000nq,Alonzo-Artiles:2021mym,Ema:2020ggo}.

Gravitational waves have been discovered by the LIGO/Virgo experiments~\cite{LIGOScientific:2016aoc} and the Einstein gravity is confirmed at the classical level. The quantum nature of the gravity, on the other hand, is not confirmed yet. 
The future discovery of quantized field of the gravitational wave, the graviton, will give us a clue to the quantum gravity.
In the universe, there are several phenomena that produce gravitons: scattering of Standard Model particles in thermal plasma~\cite{Ghiglieri:2015nfa,Ghiglieri:2020mhm,Ringwald:2020ist}, bremsstrahlung production at the inflaton decay~\cite{Nakayama:2018ptw,Huang:2019lgd} and the amplification of the vacuum fluctuation during the reheating era~\cite{Ema:2015dka,Schiappacasse:2016nei,Ema:2016hlw,Ema:2020ggo}. The inflationary production of primordial gravitational waves may also fall into this category~\cite{Maggiore:1999vm}.
There are several proposals to detect high frequency gravitons~\cite{Sabin:2014bua,Robbins:2021ucj,Ito:2019wcb,Ito:2020wxi,Aggarwal:2020olq,Berlin:2021txa}.

In this paper, we study a dark sector interacting with the SM only through gravity whose lightest particle is a scalar field $\phi$.
We show that dimension five couplings of $\phi$ to the quadratic curvature is expected for generic theories of this type
unless prohibited by symmetry, 
because they are generated when we integrate out particles heavier than $\phi$.
We systematically derive these couplings of $\phi$ to gravity based on the Schwinger-DeWitt formalism. 
In particular, we clarify that the suppression scale $\Lambda$ for these operators is associated with the mass of heavy particles and hence can be much smaller than the Planck scale.
As a result, the amount of gravitational waves is far more enhanced, 
imprinting a high frequency peak in the spectrum of gravitational waves.
Also, in some phenomenological applications, a relatively small suppression scale is required to have interesting effects, which is, however, suffering from the appearance of a ghost (see, \emph{e.g.},~\cite{Alexander:2004wk,Lyth:2005jf}).
Our understanding provides a healthy UV completion to a small suppression scale, above which a heavy particle just appears in the spectrum without having a ghost.

This paper is organized as follows. In Sec.~\ref{sec:quadratic} we systematically derive the scalar coupling to the quadratic curvature: $R^2$, $R_{\mu\nu}R^{\mu\nu}$, $R_{\mu\nu\rho\sigma}R^{\mu\nu\rho\sigma}$ and $R_{\mu\nu\rho\sigma}\widetilde R^{\mu\nu\rho\sigma}$ by integrating out heavy particles, starting from the renormalizable couplings in the matter sector. 
In Sec.~\ref{sec:scalar} we show that the scalar decay into the graviton pair occurs through the effective operator derived in Sec.~\ref{sec:quadratic}.
The graviton abundance from the scalar decay can be significant in some concrete models and we derive resultant cosmological graviton spectrum.
In Sec.~\ref{sec:summary} we summarize our results.

\section{Scalar field couplings to quadratic curvature}  \label{sec:quadratic}

\begin{figure}[t]
	\centering
 	\includegraphics[width=0.5\linewidth]{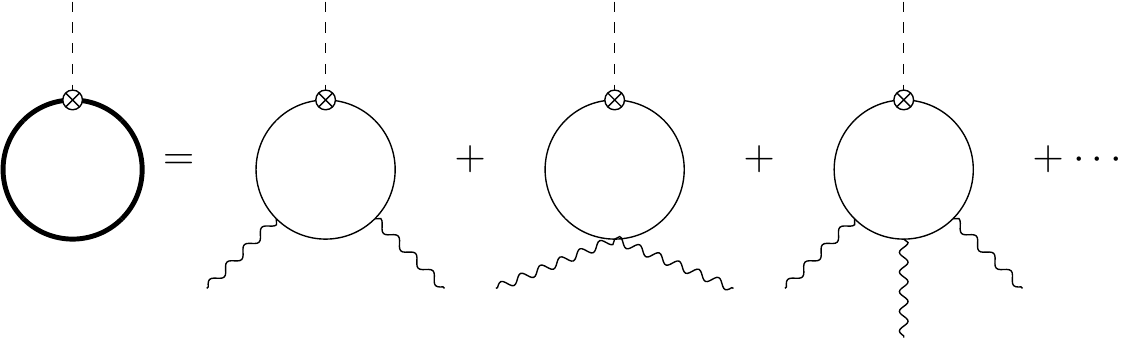}
	\caption{\small The schematic picture that the propagator induces the scalar field couplings
	to the curvature tensors. The thick line on the left hand side indicates the full propagator of the heavy
	intermediate particle,
	which contains the curvature tensors as indicated on the right hand side.
	The vertex insertion is indicated by the crossed dot, the scalar field by the dashed lines,
	and the graviton by the wavy lines.}
	\label{fig:feynman_one_loop}
\end{figure}

In this section, we derive the couplings of a scalar field to the quadratic curvature terms
induced after integrating out heavy intermediate fields.
Before going to actual computations, 
let us illustrate our basic strategy with the following simple action:
\begin{align}
	S = \int \dd^4x \sqrt{-g} \left[\frac{M_P^2}{2}R + \frac{1}{2}g^{\mu\nu}\partial_\mu \phi \partial_\nu \phi
	+ \frac{1}{2}g^{\mu\nu}\partial_\mu \chi \partial_\nu \chi 
	- \frac{m^2 + \lambda\phi}{2}\chi^2 - V(\phi) \right],
\end{align}
where $g_{\mu\nu}$ is the metric tensor with $g$ its determinant, $M_P$ is the (reduced) Planck mass,
$R$ is the Ricci scalar, $\chi$ is a heavy scalar field to be integrated out and $\phi$ is a scalar field
that eventually couples to the quadratic curvature terms.
By integrating out $\chi$, the one-loop effective action is given by
\begin{align}
	\Gamma[\phi] &= \frac{i}{2}\mathrm{Tr}\log \left[\Box + m^2 + \lambda\phi\right],
\end{align}
where $\Box = \nabla_\mu \nabla^\mu$ with $\nabla_\mu$ the covariant derivative.
Since we are interested in the term linear in $\phi$, 
we may expand the effective action as
\begin{align}
	\Gamma[\phi] &\simeq \frac{i \lambda}{2}\mathrm{Tr}\left[\frac{1}{\Box + m^2} \phi\right]
	= -\frac{\lambda}{2}\int \dd^4x \sqrt{-g} \left[D_{F}(x, x) \phi(x)\right],
	\label{eq:effective_coupling_toy}
\end{align}
where the propagator and the state are defined as
\begin{align}
	iD_F(x, x') &= \langle x \vert \frac{1}{\Box + m^2}\vert x'\rangle,
	\quad
	\langle x \vert x' \rangle = \frac{1}{\sqrt{-g}} \delta^{(4)}(x -x ').
\end{align}
Since the $\Box$-operator contains the gravitons, 
the propagator depends on the curvature tensors,
which are then translated to the scalar field couplings to the quadratic curvature terms 
through Eq.~\eqref{eq:effective_coupling_toy}.
This is schematically illustrated in Fig.~\ref{fig:feynman_one_loop}.
Thus our primary task is to compute the propagator. We consider the simple example here,
but the same strategy is applicable to a wider situation as we shall see.

The rest of this section is organized as follows.
In Sec.~\ref{subsec:propagator}, we compute the propagator of a field with a general spin.
In particular, we use the Schwinger-DeWitt formalism
and derive the propagator to the lowest order in the WKB expansion.
The WKB expansion is equivalent to the derivative expansion of the coupling,
which is suitable for our purpose since we are interested in the heavy mass limit of the fields that we integrate out.
In Sec.~\ref{subsec:integrating_out}, we compute the scalar field couplings to the quadratic curvature
terms after integrating out heavy scalar/fermion/vector fields.
Obviously the propagator that we compute in Sec.~\ref{subsec:propagator} plays an essential role there.
Finally we see that our results are related to the beta functions in Sec.~\ref{subsec:beta_fn}.

\subsection{Propagator and WKB expansion}
\label{subsec:propagator}
The propagator plays the central role in our derivation
of the scalar field couplings to the quadratic curvature terms.
In this subsection, we review the derivation of the propagator
based on the Schwinger-DeWitt formalism
(see \emph{e.g.},~\cite{DeWitt:1975ys,Birrell:1982ix,Parker:2009uva} and references therein).

We define the propagator as
\begin{align}
	\left(\Box + X \right) iD_F(x, x') = \frac{\mathbbm{1}}{\sqrt{-g}}\delta^{(d)}(x-x'),
\end{align}
where we allow $X$ dependent not only on the mass term but on the curvature tensors in general.
Here $D_F$ is understood
either scalar/bispinor/tensor depending on the spin of a particle that we integrate out,
and hence we use the bold font for the unit operator in the right hand side.
Furthermore, we anticipate the UV divergences and hence we perform the dimensional regularization.
With the proper time $\tau$ the propagator is expressed as
\begin{align}
	D_F(x, x') = \int_0^\infty \dd\tau\,K(\tau; x, x'),
	\label{eq:Green_fn_propertime}
\end{align}
where
\begin{align}
	K(\tau; x, x') &= \langle x \rvert \exp\left[-i\tau \left(\Box + X\right)\right] \lvert x'\rangle
	\equiv \langle x, \tau\vert x', 0\rangle.
\end{align}
The function $K$ satisfies the Schr\"odinger-type equation:
\begin{align}
	i\frac{\partial}{\partial \tau} K(\tau; x, x') &= \left(\Box + X \right) K(\tau; x, x'),
	\label{eq:Schrodinger}
\end{align}
and this guarantees Eq.~\eqref{eq:Green_fn_propertime},
where we use that $K(\tau; x, x') \to 0$ as $\tau \to \infty$ due to the implicit $i\epsilon$-prescription.
We now make the WKB ansatz\footnote{
	This is the so-called $R$-summed version of the WKB ansatz~\cite{Parker:1984dj,Jack:1985mw}.
}
\begin{align}
	K(\tau; x, x') = \frac{i}{\left(4\pi i\tau\right)^{d/2}} 
	\exp\left[-\frac{i\sigma(x, x')}{2\tau} - i\tau \left(X-\frac{1}{6}R\right)\right] \Delta^{1/2}(x,x')
	\Omega(\tau; x, x'),
	\label{eq:K_WKB}
\end{align}
as the solution of the Schr\"odinger equation,
where $\sigma$ is Synge's world function and $\Delta$ the van Vleck-Morette determinant (see App.~\ref{app:details}).
The dimension of spacetime is promoted to be $d$ as we use the dimensional regularization.
We expand $\Omega$ as
\begin{align}
	\Omega(\tau; x, x') = \sum_{n=0}^{\infty} \left(i\tau\right)^n a_n(x, x'),
\end{align}
with the condition $K(0; x, x') = \mathbbm{1}\delta^{(d)}(x-x')/\sqrt{-g}$ fixing
\begin{align}
	a_{0} &= \mathbbm{1}.
\end{align}
The propagator in the coincidence limit $x'\to x$ is given in terms of $a_n$ as
\begin{align}
	\lim_{x'\to x} D_F(x, x')
	&= \frac{1}{(4\pi)^{d/2}} \sum_{n=0}^{\infty}[a_n] \left(\mathcal{M}^2\right)^{d/2-n-1}
	\Gamma\left(n-\frac{d}{2}+1\right),
	\label{eq:propagator_WKB}
\end{align}
where we denote
\begin{align}
	[a_n] &\equiv \lim_{x'\to x} a_n(x,x'),
	\quad
	\mathcal{M}^2 = X - \frac{1}{6}R.
\end{align}
Remember that, for our purpose, we need the propagator only in the coincidence limit since the effective
action contains the trace of the spacetime (see Eq.~\eqref{eq:effective_coupling_toy}).
From this expression it is clear that the small $\tau$ expansion corresponds to the derivative expansion
since the higher order terms are suppressed by the higher powers of $\mathcal{M}^2$.
We can compute $[a_n]$ from the Schr\"odinger equation. We leave the derivation in App.~\ref{app:details}
and show only the final results here~\cite{Christensen:1976vb,Jack:1985mw,Avramidi:1990je}:
\begin{align}
	[a_0] &= \mathbbm{1},
	\quad
	[a_1] = 0, \label{eq:a0a1} \\
	[a_2] &=  \left[-\frac{1}{180}R_{\mu\nu}R^{\mu\nu} +\frac{1}{180} R_{\mu\nu\rho\sigma}R^{\mu\nu\rho\sigma}
	+ \frac{1}{12}W_{\mu\nu}W^{\mu\nu} + \frac{1}{6}\Box X - \frac{1}{30}\Box R\right]\mathbbm{1},
	\label{eq:a2}
\end{align}
where the field $\Psi$ that is integrated out is assumed to satisfy
\begin{align}
	\left[\nabla_\mu, \nabla_\nu\right]\Psi = W_{\mu\nu} \Psi.
\end{align}

\subsection{Integrating out heavy fields}
\label{subsec:integrating_out}
Equipped with the WKB expansion of the propagators,
we now derive the scalar field couplings to the quadratic curvature terms.
We consider the cases that the heavy intermediate particle is a scalar field, a fermion and a vector field, in turn.

\paragraph{Scalar field.}
We first integrate out a heavy scalar field. We may take the action as
\begin{align}
	S = \int \dd^4x \sqrt{-g} \left[\frac{M_P^2}{2}\left(1+\frac{\xi \chi^2}{M_P^2}\right)R 
	+ \frac{1}{2}g^{\mu\nu}\partial_\mu \phi \partial_\nu \phi
	+ \frac{1}{2}g^{\mu\nu}\partial_\mu \chi \partial_\nu \chi 
	- \frac{m^2 + \lambda\phi}{2}\chi^2 - V(\phi) \right],
\end{align}
where we now include the nonminimal coupling between $\chi$ and the Ricci scalar.
As we explained at the beginning of this section, the effective coupling is given by
\begin{align}
	S_\mathrm{eff} &= -\frac{\mu^{4-d}}{2}\int \dd^4x \sqrt{-g}\left[\lambda \phi(x) \lim_{x'\to x} D_F(x,x')\right],
\end{align}
where $\mu$ is the scale that is introduced in the dimensional regularization,
and
\begin{align}
	X = m^2 - \xi R,
	\quad
	\mathcal{M}^2 = m^2 - \left(\xi + \frac{1}{6}\right)R.
\end{align}
With Eq.~\eqref{eq:propagator_WKB}, we obtain the effective coupling to the quadratic curvature terms as
\begin{align}
	\mathcal{L}_\mathrm{eff} &= -\frac{1}{32\pi^2}\frac{\lambda \phi}{m^2}
	\left[\frac{1}{2}\left(\xi + \frac{1}{6}\right)^2 R^2 - \frac{1}{180}R_{\mu\nu}R^{\mu\nu}
	+ \frac{1}{180}R_{\mu\nu\rho\sigma}R^{\mu\nu\rho\sigma}\right],
\end{align}
where we use that $W_{\mu\nu} = 0$ for the scalar field.
In particular, if we assume that the mass of $\chi$ comes from the scalar field vacuum expectation value (VEV), 
the original interaction may be written as
\begin{align}
	\mathcal{L}_\mathrm{mass} = -\frac{\tilde{\lambda}}{2}\Phi^2 \chi^2,
\end{align}
with $\Phi = v_\phi + \phi$.
We then identify that (note that $\lambda$ is dimensionful and $\tilde{\lambda}$ is dimensionless)
\begin{align}
	m^2 = \tilde{\lambda} v_\phi^2,
	\quad
	\lambda = 2\tilde{\lambda}v_\phi,
	~~\Rightarrow~~
	\frac{\lambda}{m^2} = \frac{2}{v_\phi}.
\end{align}
We then obtain
\begin{align}
	\mathcal{L}_\mathrm{eff} = -\frac{1}{16\pi^2} \frac{\phi}{v_\phi}
	\left[\frac{1}{2}\left(\xi + \frac{1}{6}\right)^2 R^2 - \frac{1}{180}R_{\mu\nu}R^{\mu\nu}
	+ \frac{1}{180}R_{\mu\nu\rho\sigma}R^{\mu\nu\rho\sigma}\right].
\end{align}

\paragraph{Fermion.}

We next consider the fermion. We may take the action as
\begin{align}
	S = \int \dd^4x\sqrt{-g}\,\bar{\psi}\left[i\slashed{\nabla} - m -\lambda\phi\right] \psi.
\end{align}
The effective action is given by
\begin{align}
	\Gamma[\phi] &= -i \mathrm{Tr}\log\left[i \slashed{\nabla} - m - \lambda\phi\right]
	= i\lambda \mathrm{Tr}\left[\frac{1}{i\slashed{\nabla} - m}\phi\right]  + \cdots.
\end{align}
We note that the spinor traces of odd $\gamma$s vanish, and hence
\begin{align}
	\mathrm{Tr}\left[\frac{1}{i\slashed{\nabla} - m}\phi\right] 
	&= \mathrm{Tr}\left[\frac{-i\slashed{\nabla}-m}{-i\slashed{\nabla}-m}\frac{1}{i\slashed{\nabla} - m}\phi\right]
	= -m\mathrm{Tr}\left[\frac{1}{\Box + X}\phi\right],
	\quad
	X = m^2 + \frac{R}{4}.
\end{align}
We thus obtain the effective action as
\begin{align}
	S_\mathrm{eff} &=  \mu^{4-d} \int \dd^4 x\sqrt{-g}
	\left[\lambda m
	\phi(x)\,\mathrm{tr}\left[\lim_{x'\to x} D_F(x,x')\right]\right],
\end{align}
where the trace is over the spinor indices.
In the fermion case we have (see App.~\ref{app:convention})
\begin{align}
	W_{\mu\nu}W^{\mu\nu} &= -\frac{i\gamma_5}{8} R_{\mu\nu\rho\sigma} \tilde{R}^{\mu\nu\rho\sigma}
	- \frac{1}{8}R_{\mu\nu\rho\sigma} R^{\mu\nu\rho\sigma}.
\end{align}
With Eq.~\eqref{eq:propagator_WKB}, we then obtain the effective coupling as
\begin{align}
	\mathcal{L}_\mathrm{eff}
	&= \frac{1}{16\pi^2}\frac{\lambda \phi}{m}
	\left[\frac{1}{72}R^2 - \frac{1}{45}R_{\mu\nu}R^{\mu\nu} 
	- \frac{7}{360}R_{\mu\nu\rho\sigma}R^{\mu\nu\rho\sigma}\right].
\end{align}
In particular, if we assume that the mass of the fermion comes from $\phi$, 
we may take $m = \lambda v_\phi$ (here $\lambda$ is dimensionless) and we have
\begin{align}
	\mathcal{L}_\mathrm{eff}
	&= \frac{1}{16\pi^2}\frac{\phi}{v_\phi}
	\left[\frac{1}{72}R^2 - \frac{1}{45}R_{\mu\nu}R^{\mu\nu} 
	- \frac{7}{360}R_{\mu\nu\rho\sigma}R^{\mu\nu\rho\sigma}\right].
\end{align}

Here it may be appropriate to comment on the gravitational chiral anomaly.
Indeed, the coupling induced by the gravitational anomaly can be computed within the same framework.
In order to see this point, we may consider the following action:
\begin{align}
	S = \int \dd^4x\sqrt{-g}\,\bar{\psi}\left[i\slashed{\nabla} - m e^{2i\gamma_5 \phi/f}\right] \psi.
\end{align}
We may expand the effective action to linear order in $\phi$ and obtain
\begin{align}
	\Gamma[\phi] &= -i\mathrm{Tr}
	\left[\frac{1}{i\slashed{\nabla}-m}\left(-2i\gamma_5 \frac{m \phi}{f}\right)\right].
\end{align}
In the same way as above, this is rewritten as
\begin{align}
	S_\mathrm{eff} &= \mu^{4-d}\int \dd^4 x \sqrt{-g}
	\left[\frac{2i m^2 \phi}{f}\mathrm{tr}\left[\lim_{x'\to x} D_F(x,x') \gamma_5 \right]\right].
\end{align}
Only the term proportional to $\gamma_5$ in $W_{\mu\nu}W^{\mu\nu}$ contributes and we thus get
\begin{align}
	\mathcal{L}_\mathrm{eff} 
	&= \frac{1}{192\pi^2}\frac{\phi}{f}
	R_{\mu\nu\rho\sigma}\tilde{R}^{\mu\nu\rho\sigma}.
	\label{Lanomaly}
\end{align}
In order to see that this is consistent with the gravitational chiral anomaly, 
we note that the axial phase in the mass can be rotated away by
\begin{align}
	\psi \rightarrow e^{-i\gamma_5 \phi/f} \psi.
\end{align}
This rotation then induces~\cite{Fujikawa:1980eg}
\begin{align}
	\mathcal{L}_{\mathrm{anom}} &= \frac{1}{192\pi^2} \frac{\phi}{f}
	R_{\mu\nu\rho\sigma}\tilde{R}^{\mu\nu\rho\sigma},
\end{align}
through the gravitational chiral anomaly, in addition to the derivative interaction between $\phi$ and $\psi$.
Assuming that the derivative coupling does not induce further threshold corrections (which is the case
\emph{e.g.}, in the axion-gauge field system~\cite{Georgi:1986df}, and what we expect from the decoupling theorem), 
this agrees with the effective coupling that we derived.

\paragraph{Vector field.}

Finally we consider the vector boson.
To be specific, we assume that the mass originates from the Higgs mechanism and
hence consider the abelian Higgs model:
\begin{align}
	S &= \int \dd^4x\sqrt{-g}\left[-\frac{1}{4}F_{\mu\nu}F^{\mu\nu}
	+ \abs{ D_\mu \Phi}^2 - \lambda\left(\abs{\Phi}^2 - \frac{v_\phi^2}{2}\right)^2
	+\delta_{B}\left(\bar{c} F_A\right)
	\right],
\end{align}
where $A_\mu$ is the gauge field with $F_{\mu\nu}$ its field strength, 
$\delta_B$ is the BRST transformation, 
$F_A$ is the gauge fixing function which we specify below, and
\begin{align}
	D_\mu \Phi = \left(\partial_\mu + i e A_\mu\right) \Phi.
\end{align}
We expand $\Phi$ as
\begin{align}
	\Phi = \frac{v_\phi + \phi + i\chi}{\sqrt{2}}.
\end{align}
The action is then given by
\begin{align}
	\mathcal{L} &= -\frac{1}{4}F_{\mu\nu}F^{\mu\nu} 
	+ \frac{\bar{m}_A^2}{2} A^\mu A_\mu
	+ \frac{1}{2}\left(\partial \chi\right)^2 - \frac{\lambda}{2}\left(2v_\phi \phi + \phi^2\right)\chi^2 
	+ \bar{m}_A A^\mu \partial_\mu \chi
	\nonumber \\
	&+ \frac{1}{2}\left(\partial \phi\right)^2 - \frac{m_\phi^2}{2}\phi^2 - g\chi A^\mu \partial_\mu \phi
	+ \delta_B\left(\bar{c}F_A\right)
	+ \cdots,
\end{align}
where
\begin{align}
	\bar{m}_A^2 = m_A^2 \left(1+\frac{\phi}{v_\phi}\right)^2,
	\quad
	m_A^2 = e^2v_\phi^2,
	\quad
	m_\phi^2 = 2\lambda v_\phi^2,
\end{align}
and the dots indicate the higher order terms that are irrelevant at one-loop.

We now perform the derivative expansion. This is equivalent to
taking $m_\phi/m_A \to 0$ (or $e^2 \gg \lambda$) in the current case. 
Although the higher order terms in the small $m_\phi^2/m_A^2$ expansion can be obtained with the higher
order WKB expansions, here we keep only the leading order terms.
This in particular means that we should drop the $\phi$-dependent mass term of $\chi$ 
that is proportional to $\lambda$ to be consistent.
For the same reason, we should also drop the $\chi$-$A$ mixing term that depends on $\partial_\mu \phi$.
With this in mind, the quadratic action of $\chi$ and $A$ to the order of our interest is given by
\begin{align}
	\mathcal{L}_{\chi A}   
	&= -\frac{1}{4}F_{\mu\nu}F^{\mu\nu} 
	+ \frac{\bar{m}_A^2}{2} A^\mu A_\mu
	+ \frac{1}{2}\left(\partial \chi\right)^2
	+ \bar{m}_AA^\mu \partial_\mu \chi
	+ \delta_B\left(\bar{c}F_A\right).
\end{align}
We take the gauge fixing function $F_A$ as
\begin{align}
	F_A = \frac{\zeta}{2}B - \nabla_\mu A^\mu + \zeta \bar{m}_A \chi,
\end{align}
where $\zeta$ is the gauge fixing parameter.
The BRST transformation is given by
\begin{align}
	\delta_B \bar{c} = B,
	\quad
	\delta_B B = 0,
	\quad
	\delta_B A_\mu = -\partial_\mu c,
	\quad
	\delta_B \phi = -g c \chi,
	\quad
	\delta_B \chi = \bar{m}_A c.
\end{align}
By integrating out the Nakanishi-Lautrup $B$-field, we then obtain
\begin{align}
	\mathcal{L} &=-\frac{1}{4}F_{\mu\nu}F^{\mu\nu}
	- \frac{1}{2\zeta}\left(\nabla_\mu A^\mu\right)^2
	+ \frac{1}{2}\bar{m}_A^2 A_\mu A^\mu
	+ \frac{1}{2}\left(\partial \chi\right)^2
	-\frac{1}{2}\zeta \bar{m}_A^2 \chi^2
	- \bar{c}\left(\Box + \zeta \bar{m}_A^2 \right) c
	+ \cdots.
\end{align}
In the following we take the Feynman gauge $\zeta = 1$ and then we get
\begin{align}
	\mathcal{L} &= \frac{1}{2}A_\mu \left(g^{\mu\nu}\Box + g^{\mu\nu}\bar{m}_A^2 + R^{\mu\nu}\right) A_\nu
	- \frac{1}{2}\chi\left(\Box + \bar{m}_A^2\right)\chi
	- \bar{c}\left(\Box + \bar{m}_A^2 \right) c
	+ \cdots.
\end{align}
We may expand the effective action to the linear order in $\phi$ as
\begin{align}
	S_\mathrm{eff} &= - \mu^{4-d}\int \dd^4 x \sqrt{-g}
	\left[\mathrm{tr}\left[D_F^{(V)}\right] - D_F^{(S)}\right]\frac{m_A^2 \phi}{v_\phi},
\end{align}
where the superscript indicates whether a given propagator is that of a vector or scalar field,
and we use that the ghost is Grassmannian.
We note that
\begin{align}
	\left[\nabla_\mu, \nabla_\nu\right] A_\rho &= -{R_{\mu\nu\rho}}^\sigma A_\sigma
	= {\left(W_{\mu\nu}\right)_{\rho}}^{\sigma} A_\sigma,
\end{align}
and hence
\begin{align}
	\mathrm{tr}\left[W_{\mu\nu}W^{\mu\nu}\right]
	&= -R_{\mu\nu\rho\sigma}R^{\mu\nu\rho\sigma}.
\end{align}
We thus obtain
\begin{align}
	\mathcal{L}_\mathrm{eff}
	= -\frac{1}{16\pi^2}\frac{\phi}{v_\phi}
	\left[
	-\frac{1}{8}R^2 + \frac{29}{60}R_{\mu\nu}R^{\mu\nu} 
	- \frac{1}{15}R_{\mu\nu\rho\sigma}R^{\mu\nu\rho\sigma}
	\right].
\end{align}

\subsection{Relation to beta function as low energy theorem}
\label{subsec:beta_fn}

In the previous subsection we have derived the effective couplings of the scalar field
to the quadratic curvature terms.
These coefficients are actually controlled by the beta function.
Indeed, if an intermediate particle is heavy enough, its effect on the low energy theory
is only through threshold corrections.
The interaction of a scalar field $\phi$ to the heavy particle of the form $m(1+\phi/v_\phi)$ modifies 
the energy scale at which the heavy particle decouples in the constant $\phi$ limit, 
and hence the induced effective couplings are given by the beta functions.
This is analogous to the Higgs couplings to photons
in the massless Higgs limit~\cite{Shifman:1979eb}.

In the case of our interest, the relevant beta functions are given by~\cite{Salvio:2018crh}\footnote{
	The beta functions of the quadratic curvature terms are computed \emph{e.g.}, in~\cite{Salvio:2018crh,Salvio:2014soa,Salvio:2017qkx,Fradkin:1981iu,Buchbinder:1992rb,Elizalde:1993ee,
	Elizalde:1993ew,Codello:2015mba,Markkanen:2018bfx},
	although some references have opposite signs than others.
	Here we use the result in~\cite{Salvio:2018crh}
	which agrees with our computation including the sign
	and hence we believe is correct.
}
\begin{align}
	\beta_{\alpha_1} &= \frac{\dd\alpha_1}{\dd\log \mu} = \frac{1}{16\pi^2}\left(-\frac{N_S}{2}\left(\xi + \frac{1}{6}\right)^2 
	+ \frac{N_F + 20N_V}{144}\right),\\
	\beta_{\alpha_2} &= \frac{\dd\alpha_2}{\dd\log \mu} = \frac{1}{16\pi^2}\frac{N_S - 2N_F - 88N_V}{180},\\
	\beta_{\alpha_3} &= \frac{\dd\alpha_3}{\dd\log \mu} = -\frac{1}{16\pi^2}\frac{4N_S + 7N_F - 52N_V}{720},
\end{align}
where $N_S, N_F$ and $N_V$ are number of real scalars, Weyl fermions and gauge bosons, respectively.
We define the coefficients as
\begin{align}
	\mathcal{L} = \alpha_1 R^2 + \alpha_2 R_{\mu\nu}R^{\mu\nu}
	+ \alpha_3 R_{\mu\nu\rho\sigma}R^{\mu\nu\rho\sigma},
\end{align}
and drop the contribution from the graviton loops.
The RGE is different above and below the scale at which the heavy particles decouple, 
and $\phi$ modifies this scale as we explained above. Below the scale of the heavy particle, 
this threshold correction is written as
\begin{align}
	\Delta \alpha_i &= \beta_{\alpha_i}\log\left(1+\frac{\phi}{v_\phi}\right).
\end{align}
By expanding this with respect to $\phi$, we obtain the effective Lagrangian as
\begin{align}
	\mathcal{L}_\mathrm{eff}
	&= \frac{\phi}{v_\phi}\left(\beta_{\alpha_1}R^2 + \beta_{\alpha_2} R_{\mu\nu}R^{\mu\nu}
	+ \beta_{\alpha_3}R_{\mu\nu\rho\sigma}R^{\mu\nu\rho\sigma}\right).
	\label{Leff}
\end{align}
For the scalar field and fermion cases, we see that this effective Lagrangian coincides with 
our computation in Sec.~\ref{subsec:integrating_out} with $N_S = 1$ or $N_F = 2$ 
(remember that $N_F$ is the number of Weyl fermions and 
we considered the Dirac fermion in Sec.~\ref{subsec:integrating_out}).
For the vector field case, we take $N_S = N_V = 1$ with $\xi=0$ and then the above coupling coincides with
our result in Sec.~\ref{subsec:integrating_out}.
The additional scalar degree of freedom is understood as the contribution from the longitudinal mode.
Note that the same interpretation is applied to the Higgs couplings to photons 
from the $W$-boson loop in~\cite{Shifman:1979eb}.
We thus establish the relation between the effective couplings and the beta functions,
which we may call the low energy theorem following~\cite{Shifman:1979eb}.

Two comments are in order.
First, there is actually no surprise that our computation is related to the beta functions.
This is rather built-in since our method, if we do not expand with respect to $\phi$, 
is equivalent to computing the full one-loop effective action,
which after renormalization encodes the information of the beta functions.
See \emph{e.g.},~\cite{Markkanen:2018bfx} for the effective potential and 
the beta functions of the SM in the curved spacetime derived in this way.
Second, although we focused on the term linear in $\phi$ and quadratic in the curvature tensors,
the method is readily extended to higher order terms.
In particular, if we know a gravitational analogue of the (full-order) Euler-Heiseinberg Lagrangian,
we can easily derive couplings between $n$ scalar fields and $m$ curvature tensors
with $n$ and $m$ arbitrary.
Unfortunately, to our knowledge, no such an effective Lagrangian is known.
Thus, although definitely doable order by order, 
it requires an additional task to derive such an effective interaction.
This is in contrast to the Higgs couplings to photons that the Euler-Heisenberg Lagrangian
is used to derive such a generalized version of the low energy theorem~\cite{Shifman:1979eb}.

\section{Scalar decay into gravitons} \label{sec:scalar}

In this section we study a scalar field decay into a graviton pair.
First we briefly explain the lowest order operators that induce the decay.
Perhaps the simplest interaction that one may think of is $\mathcal{L} = c \phi R$.
This coupling, however, does not induce the decay, contrary to the claims
in~\cite{Zee:1978wi,Barr:1989jn,Cervantes-Cota:1995ehs}.
The key point is that $R$ contains a linear term in the (scalar part of) graviton
and hence $c$ induces a kinetic mixing between the graviton and $\phi$.
Therefore one has to solve the kinetic mixing before discussing the decay of $\phi$.
Solving the kinetic mixing is equivalent to moving to the frame without the nonminimal coupling,
or the conformal frame,
by the Weyl transformation. One does not have any coupling between $\phi$ and two gravitons 
in the conformal frame, meaning that there is no decay induced 
by the operator $\phi R$~\cite{Ema:2015dka}.
The next simplest possibility is the couplings between $\phi$ and the quadratic curvature.
Among them, $\phi R^2$ and $\phi R_{\mu\nu}R^{\mu\nu}$ 
do not induce the decay in the flat spacetime~\cite{Delbourgo:2000nq,Alonzo-Artiles:2021mym}.
This is probably most easily seen by noting that 
the on-shell graviton equation of motion gives $R=0$ and $R_{\mu\nu}=0$ in the flat spacetime.
We thus focus on the operators~\eqref{phiRR} and~\eqref{phiRtilR} in this section.

\subsection{Scalar decay rate into graviton pair}
In Sec.~\ref{sec:quadratic} we have shown that there appear scalar field couplings to the quadratic curvature in the form of $\mathcal L_{\phi RR}$~(\ref{phiRR}) and/or $\mathcal L_{\phi R\widetilde R}$~(\ref{phiRtilR}) after integrating out heavy particles.\footnote{
Note that we cannot remove these couplings by the Weyl transformation.
}
These terms induce a scalar decay into the graviton pair. Both operators give the same decay rate as\footnote{
	This rate differs from the one in~\cite{Alonzo-Artiles:2021mym} by a factor of 16.
	We believe that this difference originates from that \cite{Alonzo-Artiles:2021mym} 
	does not properly normalize the graviton field.
}
\begin{align}
	\Gamma(\phi\to 2h) = \frac{1}{4\pi} \frac{m_\phi^7}{\Lambda^2 M_{\rm Pl}^4},  \label{phi2h}
\end{align}
Given a renormalizable interaction between $\phi$ and other scalars, fermions or vector bosons, one can calculate the cutoff scale $\Lambda$, following the procedure given in the previous section. For the scalar coupling $\mathcal L_{\phi RR}$, from Eq.~(\ref{Leff}), we obtain
\begin{align}
	\Lambda = v_\phi\frac{11520\pi^2}{4N_S+7N_F-52N_V},
	\label{Lambda}
\end{align}
where for simplicity we assumed that the coupled scalar, fermion and vector boson masses are given 
in the universal form by $m(1+\phi/v_\phi)$ and $N_S, N_F$ and $N_V$ are the number of real scalar, Weyl fermion and gauge boson that are decoupled above the scale $m_\phi$.
Although the decay rate into the graviton pair (\ref{phi2h}) might be small due to the $M_{\rm Pl}^4$ suppression, it is possible that the branching fraction of $\phi$ decay into the graviton is sizable, or even dominant.

We emphasize that the scalar decay into the graviton pair is always allowed unless it is prohibited by some symmetry. For example, let us consider the action\footnote{If the VEV of $\phi$ is zero, $v_\phi$ in (\ref{Lambda}) should be regarded as $v_\phi=m/\lambda$. If, on the other hand, $m=0$ and $\phi$ has a finite VEV, $v_\phi$ in (\ref{Lambda}) is regarded as the VEV of $\phi$.}
\begin{align}
	S=\int \dd^4x\sqrt{-g}\left[ \frac{M_{\rm Pl}^2}{2}R + \frac{1}{2}g^{\mu\nu}\partial_\mu\phi\partial_\nu\phi -V(\phi)
	+ \bar{\psi}\left(i\slashed{\nabla} - m -\lambda\phi\right) \psi\right],
\end{align}
where $\phi$ does not have interaction with the Standard Model particles, \emph{i.e.}, it lives in the completely hidden sector. If the fermion $\psi$ is heavier than $\phi$, the scalar particle $\phi$ cannot decay into the fermion pair. 
However, one can integrate out the heavy fermion and obtain 
the effective operators of the form~\eqref{phiRR} and~\eqref{phiRtilR} 
as we explicitly computed in Sec.~\ref{subsec:integrating_out}, 
which causes the scalar decay into the graviton pair. Thus the scalar $\phi$ is not stable even in such a case.
Not only heavy fermion, but also scalar boson or vector boson loops also contribute to the scalar decay into the graviton pair.
Below we show a concrete example that exhibits significant graviton production through the scalar decay.

\subsection{Cosmic graviton background spectrum from scalar decay}

An example is the dark Higgs field $\Phi$ which spontaneously breaks dark gauged U(1) symmetry:
\begin{align}
	\mathcal L = \left|(\partial_\mu + ie A_\mu)\Phi \right|^2 -\frac{1}{4}F_{\mu\nu}F^{\mu\nu}-\lambda \left(|\Phi|^2-\frac{v_\phi^2}{2}\right)^2.
	\label{darkhiggs}
\end{align}
We assume that there are no direct couplings between the dark sector and the Standard Model sector.
If $e^2 > \lambda$, the radial mode of the Higgs (which is denoted by $\phi$) cannot perturbatively decay into the gauge boson pair. In such a case, $\phi$ can only decay into the graviton through the effective operator $\mathcal L_{\phi RR}$ (\ref{phiRR}). The decay rate is estimated by inserting $N_F=0$ and $N_S=N_V=1$ in~(\ref{Lambda}) as
\begin{align}
	\Gamma(\phi\to 2h)= 8\times 10^{-17}\,{\rm GeV}~\lambda\left( \frac{m_\phi}{10^{13}\,{\rm GeV}}\right)^5,
\end{align}
where the dark Higgs mass is given by $m_\phi=\sqrt{2\lambda}v_\phi$.
In terms of the cosmic temperature $T_{\rm dec}$ at which the Hubble parameter $H$ becomes equal to $\frac{1}{3}\Gamma(\phi\to 2h)$, we obtain
\begin{align}
	T_{\rm dec} \simeq 4\,{\rm GeV}~\lambda^{1/2}\left( \frac{m_\phi}{10^{13}\,{\rm GeV}}\right)^{5/2}
	\left( \frac{106.75}{g_*(T_{\rm dec})}\right)^{1/4},
\end{align}
assuming the radiation-dominated universe when $\phi$ decays, where $g_*(T)$ denotes the relativistic degrees of freedom at the temperature $T$.
Let us suppose that $H_{\rm inf} < m_\phi < m_{\rm inf}$, where $H_{\rm inf}$ and $m_{\rm inf}$ are the inflationary Hubble scale and the inflaton mass, respectively. In such a case, the dark Higgs already sits at the potential minimum and the U(1) symmetry is already broken during inflation. After inflation ends, $\phi$ particles are produced during the inflaton oscillation through the gravitational effect~\cite{Ema:2015dka}. The abundance, in terms of the energy density-to-entropy density ratio, is given by\footnote{
    Here we present a contribution from the inflaton coherent oscillation, which is also interpreted as the gravitational inflaton annihilation~\cite{Ema:2015dka,Ema:2016hlw,Ema:2018ucl,Chung:2018ayg,Mambrini:2021zpp}. The gravitational annihilation of Standard Model particles in thermal bath also produces scalar particles~\cite{Garny:2015sjg,Tang:2016vch,Tang:2017hvq,Garny:2017kha}, but it is subdominant in the most parameters we are interested in. 
}
\begin{align}
	\frac{\rho_\phi}{s}\simeq 2\times 10^{-3}\,{\rm GeV}
	\left( \frac{m_\phi}{10^{13}\,{\rm GeV}} \right)
	\left( \frac{H_{\rm inf}}{10^{12}\,{\rm GeV}} \right)
	\left( \frac{T_{\rm R}}{10^{12}\,{\rm GeV}} \right),
	\label{gravprod}
\end{align}
where $T_{\rm R}$ denotes the reheating temperature of the universe.\footnote{
	We assume that the gauge boson mass $ev_\phi$ is larger than the inflaton mass $m_{\rm inf}$ in order to suppress the gravitational gauge boson production by the inflaton oscillation~\cite{Ema:2019yrd}.
}
Thus the dark Higgs particle abundance can be substantial when it decays and the energetic gravitons are efficiently produced. Depending on the parameter choice, the dark Higgs can be dominant component of the universe before it decays and the graviton energy density becomes so large that it conflicts with the upper bound on the dark radiation energy density at the Big-bang Nucleosynthesis or the recombination epoch.
The energy density of the graviton, in terms of the density parameter $\Omega_h\equiv \rho_h/\rho_{\rm crit}$ where $\rho_h$ and $\rho_{\rm crit}$ denote the present graviton and critical energy density of the universe, is given by
\begin{align}
	\Omega_{h}&= \Omega_{\rm rad} \left( \frac{g_{*s}(T_{\rm dec})}{g_{*0}} \right) \left( \frac{g_{*s0}}{g_{*s}(T_{\rm dec})} \right)^{4/3} \frac{4}{3T_{\rm dec}}\frac{\rho_\phi}{s} \label{Omegah}\\
	&\simeq 2\times 10^{-8}\,\lambda^{-1/2}\left( \frac{10^{13}\,{\rm GeV}}{m_\phi} \right)^{3/2}
	\left( \frac{H_{\rm inf}}{10^{12}\,{\rm GeV}} \right)\left( \frac{T_{\rm R}}{10^{12}\,{\rm GeV}} \right),
\end{align}
where $\Omega_{\rm rad}$ denotes the present density parameter of the radiation, $g_{*}$ $(g_{*s})$ is the effective degrees of freedom for the energy (entropy) density, and $g_{*0}$ ($g_{*s0}$) is its present value.

Let us calculate the present-day energy spectrum of the graviton produced by the scalar decay. 
The present day graviton energy spectrum is given by
\begin{align}
	&\frac{\dd\rho_h}{\dd\ln E}=E^2\int \frac{\dd z}{H(z)} \Gamma(\phi\to 2h) n_\phi(z) a^3(z)\frac{\dd N_h}{\dd E'},\\
	&\frac{\dd N_h}{\dd E'} = 2\delta\left(E'-\frac{m_\phi}{2}\right),~~~~~~~~E'\equiv (1+z)E,
\end{align}
where $H(z)$ is the Hubble parameter at the redshift $z$. The integral is explicitly performed as
\begin{align}
	\frac{\dd\rho_h}{\dd\ln E}=\frac{16E^4}{m_\phi^4} \frac{\Gamma(\phi\to 2h) \rho_\phi(z_d)}{H(z_d)},
\end{align}
where $1+z_d\equiv m_\phi/(2E)$. The energy density $\rho_\phi$ is given by
\begin{align}
	\rho_\phi(z) = \left(\frac{\rho_\phi}{s}\right) s(z) \exp\left(-\Gamma_{\rm tot} t(z)\right),
\end{align}
where $\Gamma_{\rm tot}$ denotes the total decay width of $\phi$ and $\rho_\phi/s$ on the right-hand-side evaluated
before $\phi$ starts to decay.
If the two graviton mode is the only possible decay channel, $\Gamma_{\rm tot}=\Gamma(\phi\to 2h)$. It is convenient to plot it in terms of the density parameter
\begin{align}
	\frac{\dd \Omega_h}{\dd \ln E} \equiv \frac{1}{\rho_{\rm crit}} \frac{\dd\rho_h}{\dd\ln E},
\end{align}
where $\rho_{\rm crit}$ denotes the present critical energy density of the universe.
The peak frequency is given by
\begin{align}
	f_{\rm peak} = \frac{m_\phi}{4\pi} \frac{a(T_{\rm dec})}{a_0} \simeq 2\times 10^{22}\,{\rm Hz}~\lambda^{-1/2}\left( \frac{10^{13}\,{\rm GeV}}{m_\phi} \right)^{3/2}.
	\label{fpeak}
\end{align}

\begin{figure}[t]
  \centering
  \begin{tabular}{cc}
    \includegraphics[width=0.7\hsize]{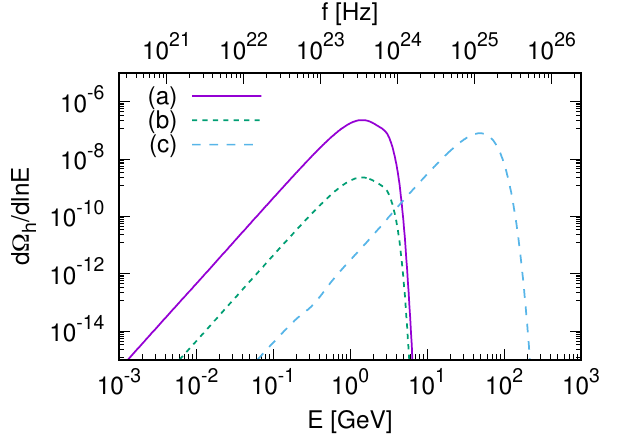}
  \end{tabular}
  \caption{The present graviton spectrum from the dark Higgs decay for (a): ($m_\phi$,$H_{\rm inf}$,$T_{\rm R}$) = ($10^{13}$,$10^{12}$,$10^{12}$), (b): $(10^{13}$,$10^{11}$,$10^{11}$)
  and (c): $(10^{12}$,$10^{11}$,$10^{11}$) in GeV unit.}
  \label{fig:spec}
\end{figure}

Fig.~\ref{fig:spec} shows the present graviton spectrum from the dark Higgs decay for (a): $(m_\phi,H_{\rm inf},T_{\rm R}) = (10^{13},10^{12},10^{12})$, (b): $(10^{13},10^{11},10^{11})$
  and (c): $(10^{12},10^{11},10^{11})$ in GeV unit.
  We have taken $v_\phi=10m_\phi$ (or $\lambda=5\times 10^{-3}$).
We used the fitting formula in~\cite{Saikawa:2018rcs} for the effective degrees of freedom for the entropy density $g_{*s}$ and energy density $g_*$ at the redshift $z_d$. A small modulation is seen in the spectrum due to the $z_d$ dependence of $g_*$ and $g_{*s}$. As seen from the figure, the typical frequency is so high that the current gravitational wave detectors are not sensitive. The reason is that the scalar decays at relatively late epoch and the effect of redshift is relatively small.

\subsection{Some other models}

Below we list several models that may predict significant amount of graviton through the scalar decay.

\paragraph{Axion.}
In the previous subsection we considered the radial mode of the dark Higgs decaying into the graviton pair. Not only the radial mode, but also the axionic component has the same decay mode. Let us consider a simple model with global chiral U(1) symmetry like the Peccei-Quinn symmetry:
\begin{align}
	S = \int \dd^4x\sqrt{-g}\,\left[ g^{\mu\nu}\partial_\mu\Phi^\dagger \partial_\nu\Phi - \lambda\left(|\Phi|^2-\frac{v_\phi^2}{2}\right)^2+ \bar{\psi}i\slashed{\nabla}\psi -(\lambda\Phi\bar{\psi}_R\psi_L+{\rm h.c.)}\right].
\end{align}
The axion-like particle $a$ appears as an angular component of $\Phi$ after the symmetry breaking: $\Phi = (v_\phi/{\sqrt 2})e^{ia/v_\phi}$. The axion obtains a mass due to either non-perturbative effects if the fermion $\psi$ are charged under some hidden gauge symmetry with strong coupling or explicit breaking terms of the global U(1).
If $\psi$ (and hidden sector gauge fields if any) is heavier than the axion, the only possible decay mode of the axion is that into the graviton pair.
The relevant effective axion-graviton coupling is given in (\ref{Lanomaly}) and the axion decay rate into the graviton pair is given by Eq.~(\ref{phi2h}) with $\Lambda=192\pi^2 v_\phi$.
The decay temperature of the axion is estimated as
\begin{align}
	T_{\rm dec} \simeq 2\times 10^{-3}\,{\rm GeV}~\left( \frac{m_a}{10^{12}\,{\rm GeV}}\right)^{7/2} \left( \frac{10^{13}\,{\rm GeV}}{v_\phi}\right)
	\left( \frac{10}{g_*(T_{\rm dec})}\right)^{1/4}.
\end{align}
The axion abundance is estimated by assuming the standard misalignment production mechanism:
\begin{align}
	\frac{\rho_a}{s}=\frac{1}{8}T_{\rm R}\left(\frac{a_i}{M_{\rm Pl}}\right)^2,  \label{rhoas}
\end{align}
where $a_i$ denotes the initial axion amplitude. Here we assumed that the axion starts to oscillate before the completion of the reheating: $m_a \gtrsim T_{\rm R}^2/M_{\rm Pl}$. Otherwise, $T_{\rm R}$ in (\ref{rhoas}) should be replaced with $\sim \sqrt{m_a M_{\rm Pl}}$.
Since the axion dominantly decays into the graviton in our setup, the graviton abundance is calculated in the same way as (\ref{Omegah}) and given by
\begin{align}
	\Omega_h \simeq 9\times 10^{-6}\left( \frac{T_{\rm R}}{10^{8}\,{\rm GeV}}\right)
	\left( \frac{v_{\phi}}{10^{13}\,{\rm GeV}}\right)^3\left( \frac{10^{12}\,{\rm GeV}}{m_a}\right)^{7/2}\left( \frac{a_i}{v_\phi}\right)^2.
\end{align}
Thus one can avoid the graviton overproduction for appropriate parameter choices, while it is also possible that a significant amount of cosmic graviton background exists in the present universe. If one would neglect the axion decay mode into the graviton, it is completely stable and it would dominate the universe leading to inconsistent cosmology. Taking account of the gravitational decay mode can make such a model viable.

\paragraph{Inflaton.}
Next we consider the possibility of the inflaton decay into the graviton pair.\footnote{
	The inflaton ``annihilation'' into the graviton pair in the Einstein gravity and its extensions has been considered in Refs.~\cite{Ema:2015dka,Ema:2016hlw,Ema:2020ggo}.
}
In this case, the relic graviton abundance is roughly given by $\Omega_h \sim \Omega_{\rm rad} \times {\rm Br}_{\phi\to 2h}$, where ${\rm Br}_{\phi\to 2h}$ denotes the branching ratio of the inflaton decay into the graviton pair. It is estimated as
\begin{align}
	{\rm Br}_{\phi\to 2h}= \frac{\Gamma(\phi\to 2h)}{\sqrt{\frac{\pi^2 g_*}{10}} \frac{T_{\rm R}^2}{M_{\rm Pl}}} \simeq 5\times 10^{-21} \left( \frac{m_{\phi}}{10^{13}\,{\rm GeV}}\right)^7 \left( \frac{10^{17}\,{\rm GeV}}{\Lambda}\right)^2
	\left(\frac{10^{10}\,{\rm GeV}}{T_{\rm R}}\right)^2. 
	\label{Br}
\end{align}
In order to avoid the constraint on the abundance of dark radiation from the cosmic microwave background observation, we need ${\rm Br}_{\phi\to 2h} \lesssim 0.1$. Note that, typical inflation models such as hilltop inflation models require nontrivial relation between the inflaton mass $m_\phi$ and its VEV $v_\phi$ (and hence $\Lambda$) once the condition for reproducing the observed large scale density perturbation is imposed. 
Under this constraint, we find that the branching ratio is typically negligibly small unless the reheating temperature is very low.

One of the possible inflation models that may predict a substantial branching ratio into the graviton pair is the so-called $\alpha$-attractor inflation model~\cite{Kallosh:2013hoa,Ferrara:2013rsa,Kallosh:2013yoa,Galante:2014ifa}. Let us modify the kinetic term of the dark Higgs in the Lagrangian (\ref{darkhiggs}) in order to identify the dark Higgs as the inflaton:
\begin{align}
	\mathcal L = \frac{\left|(\partial_\mu+ie A_\mu)\Phi \right|^2}{\left(1-\frac{|\Phi|^2}{M^2}\right)^2} -\frac{1}{4}F_{\mu\nu}F^{\mu\nu}-\lambda \left(|\Phi|^2-v_\phi^2\right)^2.
\end{align}
The potential becomes effectively flat after the canonical rescaling of the Higgs field at large field value and inflation happens there.
If we take $v_\phi < M < \sqrt{2}v_\phi$, the potential energy of the inflaton during inflation is lower than that of the origin $\Phi=0$ and hence the symmetry is never restored after inflation.
A similar inflation model was considered in Ref.~\cite{Ema:2016ops} in the context of global U(1), instead of the gauged U(1). A crucial difference is that the inflaton decays into the Goldstone boson (axion) pair in the former case while the inflaton decay to the gauge boson pair is kinematically forbidden for $e^2 > \lambda$ in the latter case.
In this model, we have a constraint on the parameter in order to reproduce the observed density perturbation of the universe~\cite{Ema:2016ops}:
\begin{align}
	\lambda\simeq 0.03 \left( \frac{10^{14}\,{\rm GeV}}{v_\phi} \right)^2 \left( \frac{x}{1-x^2} \right)^2,
\end{align}
where $x\equiv v_\phi/M$ and its range is $1/\sqrt 2 < x < 1$.
The inflaton mass is calculated as
\begin{align}
    m_\phi=2\sqrt{\lambda}v_\phi (1-x^2)\simeq 2\times 10^{13}\,{\rm GeV}\times x.
\end{align}
In this scenario, the inflaton should dominantly decay into the visible sector for successful reheating. It can be realized by introducing a small portal coupling between the Standard Model Higgs and the dark Higgs or a small kinetic mixing between the Standard Model photon and the dark photon.
The decay rate into the Standard Model sector is parametrized by the reheating temperature $T_{\rm R}$. By substituting these values into (\ref{Br}) using $\Lambda$ estimated by (\ref{Lambda}), we have ${\rm Br}_{\phi\to 2h} \sim \mathcal O(1)$ for $T_{\rm R} \sim 10\,{\rm GeV}$.

\paragraph{Decaying dark matter.}
Let us suppose that a scalar in the hidden sector is dominant component of the dark matter. 
Without a direct coupling to the Standard Model sector, it can decay into the graviton pair through the term~\eqref{phiRR} unless prohibited by symmetry.\footnote{
	Here we do not consider violation of global symmetries due to quantum gravity.
}
From the decay rate (\ref{phi2h}), the lifetime is
\begin{align}
    \tau \simeq 2\times 10^{17}\,{\rm sec}\,\left(\frac{10^{10}\,{\rm GeV}}{m_\phi}\right)^7\left(\frac{\Lambda}{M_{\rm Pl}}\right)^2.
\end{align}
Thus the lifetime can be comparable to the present age of the universe while the right amount of dark matter is obtained through the gravitational production (\ref{gravprod}) (see also Refs.~\cite{Ema:2018ucl,Ford:1986sy,Chung:1998zb,Garny:2015sjg,Tang:2016vch}). In such a case it is a simple candidate of the decaying dark matter.\footnote{
A worth-mentioning possibility is $\phi \to h h^* \to h+\mathrm{SM}$
with $h^*$ denoting an off-shell graviton and $\mathrm{SM}$ denoting
SM particles. This process involves SM particles in the final state and thus is in principle more suitable for detection. However, the decay rate is suppressed further by $m_\phi^2/M_P^2$ (and the three-body phase space factor),
and our estimation indicates that the flux from this channel is not sizable
enough to be observed by the current cosmic ray experiments.
}

\section{Summary and discussion} \label{sec:summary}

In this paper, we demonstrated that a scalar field $\phi$ can generically couple to the quadratic curvature
(unless prohibited by symmetry) 
because such couplings are generated if there exist intermediate particles heavier than $\phi$.
We systematically derived higher dimensional couplings of $\phi$ to gravity by integrating out heavy scalar, fermion, and vector fields, based on the Schwinger-DeWitt formalism at the lowest order in the WKB expansion.
We confirmed that our results are consistent with the beta functions of the quadratic curvature~\cite{Salvio:2018crh}
since the effects of heavy fields should be imprinted in threshold corrections.

One of the most interesting consequences is a decay of $\phi$ into graviton pairs, which is induced by the quadratic curvature in the form of $\mathcal L_{\phi RR}$ (\ref{phiRR}) and/or $\mathcal L_{\phi R\widetilde R}$ (\ref{phiRtilR}).
Let us note a typical frequency range of the present-day graviton spectrum from various sources. 
The dark Higgs decay as studied in the main text produces the peak graviton frequency of $f_{\rm peak}\sim 10^{22}\,{\rm Hz}$ as shown in Eq.~(\ref{fpeak}) and Fig.~\ref{fig:spec}. In the case of decaying dark matter, even higher frequency is expected: $f_{\rm peak}\sim 10^{31}\,{\rm Hz}$. On the other hand, the graviton spectrum from the Standard Model thermal plasma predicts $f_{\rm peak}\sim 10^{11}\,{\rm Hz}$~\cite{Ghiglieri:2015nfa,Ringwald:2020ist}.
The inflaton annihilation into the graviton pair also predicts the spectrum that extends to $f\sim 10^{12}\,{\rm Hz}$~\cite{Ema:2020ggo}. It shows that there are quite rich sources of high-frequency gravitons that can be probes of the early universe and high energy physics.

As mentioned in the introduction, there are several phenomenological implications of the couplings between $\phi$ 
and the quadratic curvature other than the graviton production.
For instance, the Chern-Simons coupling, $\phi R_{\mu\nu\rho\sigma} \tilde R^{\mu\nu\rho\sigma}$, can lead to gravitational leptogenesis if we identify $\phi$ as inflaton~\cite{Alexander:2004us}.
This coupling induces the production of chiral gravitational waves during inflation, which results in the $B-L$ asymmetry through the gravitational anomaly.
However, it is known that $\phi R_{\mu\nu\rho\sigma} \tilde R^{\mu\nu\rho\sigma}$ leads to the ghost instability,
which severely constrains the efficiency of leptogenesis~\cite{Alexander:2004wk,Lyth:2005jf}.
In connection with this, the prediction is highly sensitive to the UV physics and hence depends on other higher dimensional operators involving $\phi$ and curvature~\cite{Kamada:2020jaf}.
Our results suggest that such ghost instability can be absent in a healthy UV completion, where a new fermion appears above the instability scale without having a ghost.
Also, the coefficients of relevant higher dimensional operators are fixed when we integrate out a heavy fermion.
It would be interesting to study the production of chiral gravitational waves~\cite{Adshead:2018oaa} in a concrete healthy UV completion.
The same coupling is also discussed in the context of topological superconductor~\cite{Ryu2012,Wang:2010xh,Shiozaki:2013wda}.

Our results are readily extended to higher order terms both in $\phi$ and curvature.
In the context of Schwinger effect, it is known that the Schwinger current has a contribution which is only suppressed by a power of mass although the particle production should exhibit an exponential suppression for heavy masses.
This contribution can be elegantly understood as higher dimensional operators obtained from integrating out heavy particles, \emph{i.e.}, the Euler-Heisenberg Lagrangian~\cite{Banyeres:2018aax,Domcke:2019qmm}.
Similar power-suppressed behavior is also observed in the calculation of two point functions for heavy particles, $m^2 > R$, in the curved spacetime~\cite{Parker:2009uva}.
We expect that they are not related to the production of particles, rather should be identified with higher dimensional operators, gravitational analogue of the Euler-Heisenberg Lagrangian.
We will come back to this issue elsewhere.


\section*{Acknowledgments}
YE is supported in part by U.S. Department of Energy Grant No.\ desc0011842.
KM was supported by MEXT Leading Initiative for Excellent Young Researchers Grant No.\ JPMXS0320200430.
This work was supported by JSPS KAKENHI Grant Nos.\ 18K03609 (KN) and 17H06359 (KN).
The Feynman diagrams are drawn with \texttt{TikZ-Feynman}~\cite{Ellis:2016jkw}.

\appendix

\section{Convention}
\label{app:convention}

In this appendix, we summarize our conventions.
We work with the mostly-minus convention for the spacetime metric.
In particular, the Minkowski metric is given by
\begin{align}
	\eta_{ab} = \mathrm{diag}\left(+1, -1, -1, -1\right).
\end{align}
We define the geometrical quantities as
\begin{align}
	{\Gamma^{\mu}}_{\nu\rho} &= \frac{1}{2}g^{\mu\alpha}
	\left(\partial_\nu g_{\rho \alpha} + \partial_\rho g_{\nu \alpha} - \partial_\alpha g_{\nu\rho}\right), \\
	{R_{\mu\nu\rho}}^{\sigma} &= \partial_\mu {\Gamma^{\sigma}}_{\nu\rho} 
	- \partial_{\nu} {\Gamma^{\sigma}}_{\mu\rho}
	+ {\Gamma^{\alpha}}_{\nu\rho}{\Gamma^{\sigma}}_{\alpha\mu} 
	- {\Gamma^{\alpha}}_{\mu\rho}{\Gamma^{\sigma}}_{\alpha\nu}, \\
	R_{\mu\nu} &= {R_{\mu\alpha\nu}}^\alpha = \partial_\mu {\Gamma^{\alpha}}_{\alpha \nu} 
	- \partial_{\alpha} {\Gamma^{\alpha}}_{\mu\nu}
	+ {\Gamma^{\alpha}}_{\beta \nu} {\Gamma^{\beta}}_{\alpha\mu}
	- {\Gamma^{\alpha}}_{\mu\nu}{\Gamma^{\beta}}_{\alpha\beta}, \\
	R &= g^{\mu\nu}R_{\mu\nu}.
\end{align}
This fixes the sign convention for the Ricci scalar. In particular, the Einstein-Hilbert action comes with plus,
and the conformal coupling corresponds to $\xi = -1/6$ with this convention.
The covariant derivative acts on a vector field as
\begin{align}
	\nabla_\mu v_\nu = \partial_\mu v_\nu - \Gamma_{\mu\nu}^{\rho} v_\rho.
\end{align}
The commutator is then given by
\begin{align}
	\left[\nabla_\mu, \nabla_\nu\right]v_{\rho} = - {R_{\mu\nu\rho}}^\sigma v_{\sigma}.
\end{align}
We thus obtain
\begin{align}
	\left(W_{\mu\nu}W^{\mu\nu}\right)_{\alpha\beta}
	&= R_{\mu\nu\alpha\rho} {R^{\mu\nu\rho}}_\beta,
\end{align}
for a vector field.
The covariant derivative acting on a spinor is given by
\begin{align}
	\nabla_\mu \psi &= \left(\partial_\mu + \frac{1}{4}\omega_{\mu}^{ab}\gamma_{ab}\right)\psi,
\end{align}
where our convention of the gamma matrix is
\begin{align}
	\left\{\gamma^a, \gamma^b\right\} &= 2\eta^{ab},
	\quad
	\gamma_{ab} = \frac{1}{2}\left(\gamma_a \gamma_b - \gamma_b \gamma_a\right),
	\quad
	\gamma_5 = i\gamma^0 \gamma^1 \gamma^2 \gamma^3
	= - \frac{i}{4!}\epsilon^{abcd}\gamma_a \gamma_b \gamma_c\gamma_d,
\end{align}
with $\epsilon^{0123} = +1$,
and we use $a, b, \cdots$ for the local Lorentz indices and $\mu, \nu, \cdots$ for the spacetime indices.
The spin connection is defined as
\begin{align}
	\omega_{\mu}^{ab} &= e^a_\nu \left[\partial_\mu e^{\nu b} + {\Gamma^\nu}_{\sigma\mu} e^{\sigma b}\right],
\end{align}
with the vierbein defined as
\begin{align}
	g_{\mu\nu} = e_{\mu}^a \eta_{ab} e_{\nu}^b.
\end{align}
Note that this is equivalent to the compatibility equation:
\begin{align}
	\nabla_\mu e_{\nu}^a = \partial_\mu e_{\nu}^a + {\omega_{\mu}}^{ab} e_{\nu b} 
	- {\Gamma^{\rho}}_{\mu\nu}e_\rho^a = 0.
\end{align}
The Riemann tensor is equally expressed by the spin connection as
\begin{align}
	{R_{\mu\nu}}^{ab} = -\partial_\mu {\omega_{\nu}}^{ab} + \partial_\nu {\omega_{\mu}}^{ab}
	-{\omega_{\mu}}^{ac}{\omega_{\nu c}}^{b} + {\omega_\nu}^{ac} {\omega_{\mu c}}^b.
\end{align}
It indeed satisfies
\begin{align}
	R_{\mu\nu\rho\sigma} = e_{\rho a} e_{\sigma b} {R_{\mu\nu}}^{ab},
\end{align}
which one can show from $\left[\nabla_\mu, \nabla_\nu\right] e^a_\rho = 0$.
It follows that the spinor satisfies
\begin{align}
	\left[\nabla_\mu, \nabla_\nu\right] \psi = -\frac{1}{4}{R_{\mu\nu}}^{ab}\gamma_{ab} \psi.
\end{align}
We thus have
\begin{align}
	W_{\mu\nu}W^{\mu\nu} &= -\frac{i\gamma_5}{8} R_{\mu\nu\rho\sigma} \tilde{R}^{\mu\nu\rho\sigma}
	- \frac{1}{8}R_{\mu\nu\rho\sigma} R^{\mu\nu\rho\sigma},
\end{align}
where the dual Riemann tensor is defined as
\begin{align}
	\tilde{R}_{\mu\nu\rho\sigma} &= \frac{1}{2}{\epsilon_{\mu\nu}}^{\alpha\beta} R_{\alpha\beta\rho\sigma}.
\end{align}
We can also easily show that
\begin{align}
	\slashed{\nabla}^2 = \Box+ \frac{R}{4}.
\end{align}

\section{Technical details of the WKB expansion}
\label{app:details}
In this appendix we outline the computation of the coefficients of the WKB expansion $a_n$ 
in the coincidence limit. Our goal is to derive Eqs.~\eqref{eq:a0a1} and~\eqref{eq:a2}.

\subsection{Bitensor}
\label{app:bitensor}
We first summarize basic properties of Synge's world function and the related bitensors 
that are needed for the WKB approximation. We mainly follow~\cite{Poisson:2011nh}.

\paragraph{Synge's world function.}
The world function $\sigma(x, x')$ is defined as
\begin{align}
	\sigma(x, x') = \frac{1}{2}\left(\lambda_1 - \lambda_0\right) \int_{\lambda_0}^{\lambda_1} d\lambda\, g_{\mu\nu}(z(\lambda)) 
	\frac{dz^\mu}{d\lambda} \frac{dz^\nu}{d\lambda},
\end{align}
where $z(\lambda)$ is the geodesic that connects $x$ and $x'$
and $\lambda$ the affine parameter with $z(\lambda_1) = x$ and $z(\lambda_0) = x'$.
We assume that there is only one such geodesic that is true in the coincidence limit $x' \rightarrow x$.
One can show with the geodesic equation that
\begin{align}
	\epsilon \equiv g_{\mu\nu} \frac{dz^\mu}{d\lambda} \frac{dz^\nu}{d\lambda},
\end{align}
is constant along the geodesic, and hence
\begin{align}
	\sigma(x, x') = \frac{\epsilon}{2}(\lambda_1 - \lambda_0)^2 = \frac{s^2}{2},
\end{align}
where $s$ is the proper distance that is well-defined if $x$ and $x'$ are time like.

We now consider the derivative of $\sigma$ with respect to $x$ and $x'$. We may use the notation that
\begin{align}
	\sigma_{\alpha} = \frac{\partial \sigma}{\partial x^\alpha},
	\quad
	\sigma_{\alpha'}  = \frac{\partial \sigma}{\partial x^{\alpha'}},
\end{align}
and so on. We first take $x^\mu$ to $x^\mu + \delta x^\mu$ with $\delta x^\mu$ infinitesimal.
In this case the geodesic is modified as $z^\mu(\lambda) \rightarrow z^\mu(\lambda) + \delta z^\mu(\lambda)$ 
with $\delta z^\mu(\lambda_0) = 0$ and $\delta z^\mu(\lambda_1) = \delta x^\mu$.
Note that the affine parameter is rescaled such that it again runs from $\lambda_0$ to $\lambda_1$.
One can show with the geodesic equation that the variance of $\sigma$ comes only from the boundary,
and is explicitly given as
\begin{align}
	\delta \sigma = \left(\lambda_1 - \lambda_0\right)\left[g_{\mu\nu}\frac{dz^\mu}{d\lambda} \delta z^\nu\right]_{\lambda_0}^{\lambda_1}.
\end{align}
This indicates that
\begin{align}
	\sigma_\alpha = \left(\lambda_1 - \lambda_0\right) g_{\alpha \mu} t^\mu,
	\quad
	\sigma_{\alpha'} = -\left(\lambda_1 - \lambda_0\right) g_{\alpha' \mu'} {t'}^\mu,
\end{align}
where $g_{\alpha \mu}$ is the metric tensor at $x$ and $g_{\alpha' \mu'}$ at $x'$, and
\begin{align}
	t^\mu = \left.\frac{dz^\mu}{d\lambda}\right\rvert_{\lambda = \lambda_1},
	\quad
	{t'}^{\mu} = \left.\frac{dz^\mu}{d\lambda}\right\rvert_{\lambda = \lambda_0}.
\end{align}
We then obtain
\begin{align}
	\sigma^\alpha \sigma_\alpha 
	&= \sigma^{\alpha'} \sigma_{\alpha'} = 2\sigma.
	\label{eq:sigmasigma_sigma}
\end{align}

\paragraph{van Vleck-Morette determinant.}
Next we introduce the van Vleck-Morette determinant.
For this purpose we first introduce the metric bitensor $g_{\mu\nu'}$.
In general, the metric tensor is expressed by the vierbein as
\begin{align}
	g_{\mu\nu} = \eta_{ab} e^{a}_\mu e^{b}_\nu.
\end{align}
We introduce a function $g_{\mu\nu'}$ that depends both on $x$ and $x'$ as
\begin{align}
	g_{\mu\nu'}(x, x') = \eta_{ab}e^{a}_\mu(x) e^{b}_{\nu'}(x').
\end{align}
We then define
\begin{align}
	{\Delta^{\alpha'}}_{\beta'} = - {g^{\alpha'}}_\alpha {\sigma^{\alpha}}_{\beta'},
	\quad
	\Delta = \mathrm{det}{\Delta^{\alpha'}}_{\beta'}.
\end{align}
By using that $\mathrm{det}e^a_\mu = \sqrt{-g}$, this is equally expressed as
\begin{align}
	\Delta = - \frac{\mathrm{det}\left(-\sigma_{\alpha \beta'}\right)}{\sqrt{-g}\sqrt{-g'}}.
\end{align}

We now show $(\Delta \sigma^\alpha)_{;\alpha} = d\Delta$. We take the derivative of $\sigma = \sigma^\mu \sigma_\mu/2$ twice and get
\begin{align}
	\sigma_{\alpha\beta'} = {\sigma^\mu}_\alpha \sigma_{\mu\beta'} + \sigma^\mu \sigma_{\mu\alpha \beta'}.
\end{align}
The primed and unprimed derivatives commute with each other. Moreover, since $\sigma$ is a scalar function, we have 
$\sigma_{\mu\nu} = \sigma_{\nu\mu}$. As a result we obtain
\begin{align}
	{\Delta^{\alpha'}}_{\beta'} = {g^{\alpha'}}_\alpha g_{\mu\nu'} \sigma^{\mu\alpha} {\Delta^{\nu'}}_{\beta'}
	+ \sigma^\mu \left({\Delta^{\alpha'}}_{\beta'}\right)_{;\mu}.
\end{align}
By multiplying ${(\Delta^{-1})^{\beta'}}_{\gamma'}$ and taking the trace of $\alpha'$ and $\gamma'$, 
we obtain
\begin{align}
	d&= {\sigma^\alpha}_\alpha + \sigma^\alpha \left(\ln \Delta\right)_{; \alpha},
\end{align}
which is equivalent to $(\Delta \sigma^\alpha)_{;\alpha} = d\Delta$, or
\begin{align}
	d \Delta^{1/2} = \Delta^{1/2} {\sigma^{\alpha}}_\alpha + 2 \sigma^{\alpha} (\Delta^{1/2})_{\alpha}.
	\label{eq:delDelta_Delta}
\end{align}

\paragraph{Coincidence limit.}
Our primary interest is $\sigma$ and $\Delta$ in the coincidence limit $x' \rightarrow x$.
We assume that this limit exits independent of the direction that $x'$ approaches to $x$.
For our purpose its enough to consider the unprimed derivatives and hence we focus on the unprimed indexed quantities.
We first note that
\begin{align}
	[\sigma] = [\sigma_\alpha] = 0.
\end{align}
Eq.~\eqref{eq:sigmasigma_sigma} is our basic tool to study the higher derivative terms.
We take the derivative of Eq.~\eqref{eq:sigmasigma_sigma} and obtain
\begin{align}
	\sigma^{\alpha}(\sigma_{\alpha\beta} - g_{\alpha\beta}) = 0.
\end{align}
Since $\sigma^{\alpha} \propto t^{\alpha}$ and we assume that the limit is independent of $t^{\alpha}$,
we obtain
\begin{align}
	[\sigma_{\alpha\beta}] = g_{\alpha\beta}.
\end{align}
We next take the derivative of Eq.~\eqref{eq:sigmasigma_sigma} three times and obtain
\begin{align}
	[\sigma_{\gamma \alpha \beta}] + [\sigma_{\beta\alpha\gamma}] = 0.
\end{align}
The first two indices of $\sigma_{\alpha \beta\gamma}$ commute with each other
and hence this is equivalent to
\begin{align}
	2[\sigma_{\alpha \beta \gamma}] + [\left[\nabla_\beta, \nabla_\gamma\right]\sigma_{\alpha}] = 
	2[\sigma_{\alpha \beta \gamma}] - [ {R_{\beta\gamma\alpha}}^{\delta}\sigma_{\delta}] = 
	0.
\end{align}
The curvature is well-behaved in the coincidence limit,
and hence we obtain
\begin{align}
	[\sigma_{\alpha \beta\gamma}] = 0.
\end{align}
In the same way, we can show that
\begin{align}
	[\sigma_{\alpha\beta\gamma\delta}] &= S_{\alpha\beta\gamma\delta} \equiv 
	\frac{1}{3}\left(R_{\alpha\gamma\beta\delta} + R_{\alpha\delta\beta\gamma} \right), \\
	[\sigma_{\alpha\beta\gamma\delta\mu}] &= \frac{3}{4}\left[\nabla_\mu S_{\alpha\beta\gamma\delta}
	+ \nabla_\delta S_{\alpha\beta\gamma\mu} + \nabla_\gamma S_{\alpha\beta\delta\mu}\right], \\
	[\Box^3 \sigma] &= 
	\frac{4}{15}\left(R_{\mu\nu}R^{\mu\nu} - R_{\mu\nu\rho\sigma}R^{\mu\nu\rho\sigma}\right) + \frac{8}{5}\Box R.
\end{align}

We now move to the coincidence limit of the determinant. It is trivial to see that\footnote{
	Here we actually use $[\sigma_{\alpha\beta'}] = - g_{\alpha\beta}$ where the minus sign originates
	from the opposite sign of $[\sigma_\alpha]$ and $[\sigma_{\alpha'}]$.
}
\begin{align}
	[\Delta^{1/2}] = 1.
\end{align}
In order to compute the coincidence limit with derivatives, we use Eq.~\eqref{eq:delDelta_Delta}.
By taking the derivative of this equation once, we obtain
\begin{align}
	[(\Delta^{1/2})_{\alpha}] = 0.
\end{align}
By taking the derivative twice we obtain\footnote{
	It seems both the Riemann tensor and the Ricci tensor have the opposite sign as~\cite{Christensen:1976vb}.
	In our case it correctly reproduces the factor $\xi + 1/6$ so should be consistent.
}
\begin{align}
	[(\Delta^{1/2})_{\alpha\beta}] = - \frac{1}{6}R_{\alpha\beta}.
\end{align}
In the same way, we obtain
\begin{align}
	[(\Delta^{1/2})_{\alpha\beta\gamma}] &= 
	-\frac{1}{12}\left[\nabla_\gamma R_{\alpha\beta} + \nabla_\beta R_{\gamma\alpha}
	+ \nabla_\alpha R_{\beta\gamma}\right], \\
	[\Box^2 (\Delta^{1/2})] 
	&= \frac{1}{30}\left(-R_{\mu\nu}R^{\mu\nu} + R_{\mu\nu\rho\sigma}R^{\mu\nu\rho\sigma}\right)
	 - \frac{1}{5}\Box R + \frac{1}{36}R^2.
\end{align}

\subsection{WKB expansion}
\label{subsec:WKB_general}
We now compute the coefficients of the WKB expansion $a_n$ in the coincidence limit.
By substituting Eq.~\eqref{eq:K_WKB} to the Schr\"odinger equation, we obtain
\begin{align}
	0 &= i\frac{\partial \Omega}{\partial \tau} + \frac{i}{\tau}\sigma^\alpha \Omega_\alpha
	- \Delta^{-1/2} \Box \left(\Delta^{1/2} \Omega\right)
	\nonumber \\
	&+ 2i\tau (\mathcal{M}^2)^\alpha \Delta^{-1/2} \nabla_\alpha\left(\Delta^{1/2} \Omega\right)
	+ \left[\sigma^\alpha (\mathcal{M}^2)_\alpha- \frac{R}{6} + i\tau \Box \tilde{X} 
	- (i\tau)^2 (\mathcal{M}^2)^\alpha (\mathcal{M}^2)_\alpha\right]\Omega.
\end{align}
By expanding $\Omega$ as
\begin{align}
	\Omega(\tau; x, x') = \sum_{n=0}^{\infty} \left(i\tau\right)^n a_n(x, x'),
\end{align}
we obtain the recursion relations
\begin{align}
	\mathbbm{1} &= a_{0}, \\
	0 &= \sigma^{\alpha}a_{0;\alpha}, \label{eq:a0_generator}\\
	0 &= a_1 + \sigma^\alpha a_{1;\alpha} + \Delta^{-1/2} \Box \left(\Delta^{1/2} a_0\right) 
	- \left(\sigma^{\alpha}(\mathcal{M}^2)_\alpha - \frac{R}{6}\right)  a_0, \\
	0 &= 2a_2 + \sigma^\alpha a_{2;\alpha} + \Delta^{-1/2} \Box \left(\Delta^{1/2} a_1\right) 
	\nonumber \\
	&- \left(\sigma^{\alpha}(\mathcal{M}^2)_\alpha - \frac{R}{6}\right) a_1
	- 2(\mathcal{M}^2)^\alpha \Delta^{-1/2} \nabla_\alpha\left(\Delta^{1/2}a_0\right) 
	- \left(\Box \mathcal{M}^2\right) a_0, \\
	0&= \left(n+3\right) a_{n+3} + \sigma^\alpha a_{n+3;\alpha} + \Delta^{-1/2}\Box \left(\Delta^{1/2} a_{n+2}\right)
	-\left(\sigma^{\alpha}(\mathcal{M}^2)_\alpha - \frac{R}{6}\right)  a_{n+2} 
	\nonumber \\
	&- 2(\mathcal{M}^2)^\alpha \Delta^{-1/2} \nabla_\alpha\left(\Delta^{1/2}a_{n+1}\right) - 
	\left(\Box \mathcal{M}^2\right) a_{n+1}
	+(\mathcal{M}^2)^\alpha (\mathcal{M}^2)_\alpha a_{n},
	~~n \geq 0.
\end{align}
Notice that we used the bold font for $a_0$ since it lives in the Lorentz tensor space in general.

We now take the coincidence limit.
We need up to four derivatives of $a_0$ to compute the terms of our interest.
We first take the derivative of Eq.~\eqref{eq:a0_generator} once and obtain
\begin{align}
	[a_{0;\alpha}] &= 0.
\end{align}
Next, we take the derivative of Eq.~\eqref{eq:a0_generator} twice and obtain
\begin{align}
	[a_{0;\alpha\beta}] + [a_{0;\beta\alpha}] = 0.
	\label{eq:spin_anticommutator}
\end{align}
We also note that, if a particle of one's interest has a finite spin, its covariant derivative acting on $a_0$ satisfies
\begin{align}
	\left[\nabla_\alpha, \nabla_\beta\right] a_0 = W_{\alpha\beta} a_0,
\end{align}
where $W_{\alpha\beta}$ depends on the spin and is a combination of the curvature tensors.
This together with Eq.~\eqref{eq:spin_anticommutator} tell us that
\begin{align}
		[a_{0;\alpha\beta}] &= -\frac{1}{2}W_{\alpha\beta} \mathbbm{1}.
\end{align}
We then take the derivative of Eq.~\eqref{eq:a0_generator} three times and obtain
\begin{align}
	[a_{0;\alpha\beta\gamma}] + [a_{0;\beta\alpha\gamma}] + [a_{0;\gamma\alpha\beta}] = 0.
\end{align}
We have
\begin{align}
	a_{0;\beta\alpha\gamma} &= 
	a_{0;\alpha\beta\gamma} + \nabla_\gamma\left(W_{\alpha\beta} a_0\right), \\
	a_{0;\alpha\gamma\beta} &=
	a_{0;\alpha\beta\gamma} - {R_{\beta\gamma\alpha}}^{\rho}a_{0;\rho} + W_{\beta\gamma}a_{0;\alpha},
\end{align}
and hence we obtain
\begin{align}
	[a_{0;\alpha\beta\gamma}] &= 
	-\frac{1}{3}\left(W_{\alpha\beta;\gamma} + W_{\alpha\gamma;\beta}\right)\mathbbm{1}.
\end{align}
We finally take the derivative of Eq.~\eqref{eq:a0_generator} four times.
After a similar (but longer) manipulation as above we obtain
\begin{align}
	\left[\Box^2 a_0\right] = \frac{1}{2}W_{\mu\nu}W^{\mu\nu}.
\end{align}
By substituting these results in the recursion relations, we obtain
\begin{align}
	[a_0] &= \mathbbm{1},
	\quad
	[a_1] = 0, \\
	[a_2] &=  \left[-\frac{1}{180}R_{\mu\nu}R^{\mu\nu} +\frac{1}{180} R_{\mu\nu\rho\sigma}R^{\mu\nu\rho\sigma}
	+ \frac{1}{12}W_{\mu\nu}W^{\mu\nu} + \frac{1}{6}\Box X - \frac{1}{30}\Box R\right]\mathbbm{1},
\end{align}
in agreement with~\cite{Christensen:1976vb,Jack:1985mw,Avramidi:1990je}.

\small
\bibliographystyle{utphys}
\bibliography{ref}

\providecommand{\href}[2]{#2}\begingroup\raggedright\begin{thebibliography}{10}

\bibitem{Ema:2015dka}
Y.~Ema, R.~Jinno, K.~Mukaida, and K.~Nakayama, ``{Gravitational Effects on
  Inflaton Decay},''
  \href{http://dx.doi.org/10.1088/1475-7516/2015/05/038}{{\em JCAP} {\bfseries
  05} (2015) 038}, \href{http://arxiv.org/abs/1502.02475}{{\ttfamily
  arXiv:1502.02475 [hep-ph]}}.

\bibitem{Garny:2015sjg}
M.~Garny, M.~Sandora, and M.~S. Sloth, ``{Planckian Interacting Massive
  Particles as Dark Matter},''
  \href{http://dx.doi.org/10.1103/PhysRevLett.116.101302}{{\em Phys. Rev.
  Lett.} {\bfseries 116} no.~10, (2016) 101302},
  \href{http://arxiv.org/abs/1511.03278}{{\ttfamily arXiv:1511.03278
  [hep-ph]}}.

\bibitem{Markkanen:2015xuw}
T.~Markkanen and S.~Nurmi, ``{Dark matter from gravitational particle
  production at reheating},''
  \href{http://dx.doi.org/10.1088/1475-7516/2017/02/008}{{\em JCAP} {\bfseries
  02} (2017) 008}, \href{http://arxiv.org/abs/1512.07288}{{\ttfamily
  arXiv:1512.07288 [astro-ph.CO]}}.

\bibitem{Ema:2016hlw}
Y.~Ema, R.~Jinno, K.~Mukaida, and K.~Nakayama, ``{Gravitational particle
  production in oscillating backgrounds and its cosmological implications},''
  \href{http://dx.doi.org/10.1103/PhysRevD.94.063517}{{\em Phys. Rev. D}
  {\bfseries 94} no.~6, (2016) 063517},
  \href{http://arxiv.org/abs/1604.08898}{{\ttfamily arXiv:1604.08898
  [hep-ph]}}.

\bibitem{Schiappacasse:2016nei}
E.~D. Schiappacasse and L.~H. Ford, ``{Graviton Creation by Small Scale Factor
  Oscillations in an Expanding Universe},''
  \href{http://dx.doi.org/10.1103/PhysRevD.94.084030}{{\em Phys. Rev. D}
  {\bfseries 94} no.~8, (2016) 084030},
  \href{http://arxiv.org/abs/1602.08416}{{\ttfamily arXiv:1602.08416 [gr-qc]}}.

\bibitem{Tang:2016vch}
Y.~Tang and Y.-L. Wu, ``{Pure Gravitational Dark Matter, Its Mass and
  Signatures},'' \href{http://dx.doi.org/10.1016/j.physletb.2016.05.045}{{\em
  Phys. Lett. B} {\bfseries 758} (2016) 402--406},
  \href{http://arxiv.org/abs/1604.04701}{{\ttfamily arXiv:1604.04701
  [hep-ph]}}.

\bibitem{Tang:2017hvq}
Y.~Tang and Y.-L. Wu, ``{On Thermal Gravitational Contribution to Particle
  Production and Dark Matter},''
  \href{http://dx.doi.org/10.1016/j.physletb.2017.10.034}{{\em Phys. Lett. B}
  {\bfseries 774} (2017) 676--681},
  \href{http://arxiv.org/abs/1708.05138}{{\ttfamily arXiv:1708.05138
  [hep-ph]}}.

\bibitem{Garny:2017kha}
M.~Garny, A.~Palessandro, M.~Sandora, and M.~S. Sloth, ``{Theory and
  Phenomenology of Planckian Interacting Massive Particles as Dark Matter},''
  \href{http://dx.doi.org/10.1088/1475-7516/2018/02/027}{{\em JCAP} {\bfseries
  02} (2018) 027}, \href{http://arxiv.org/abs/1709.09688}{{\ttfamily
  arXiv:1709.09688 [hep-ph]}}.

\bibitem{Ema:2018ucl}
Y.~Ema, K.~Nakayama, and Y.~Tang, ``{Production of Purely Gravitational Dark
  Matter},'' \href{http://dx.doi.org/10.1007/JHEP09(2018)135}{{\em JHEP}
  {\bfseries 09} (2018) 135}, \href{http://arxiv.org/abs/1804.07471}{{\ttfamily
  arXiv:1804.07471 [hep-ph]}}.

\bibitem{Garny:2018grs}
M.~Garny, A.~Palessandro, M.~Sandora, and M.~S. Sloth, ``{Charged Planckian
  Interacting Dark Matter},''
  \href{http://dx.doi.org/10.1088/1475-7516/2019/01/021}{{\em JCAP} {\bfseries
  01} (2019) 021}, \href{http://arxiv.org/abs/1810.01428}{{\ttfamily
  arXiv:1810.01428 [hep-ph]}}.

\bibitem{Chung:2018ayg}
D.~J.~H. Chung, E.~W. Kolb, and A.~J. Long, ``{Gravitational production of
  super-Hubble-mass particles: an analytic approach},''
  \href{http://dx.doi.org/10.1007/JHEP01(2019)189}{{\em JHEP} {\bfseries 01}
  (2019) 189}, \href{http://arxiv.org/abs/1812.00211}{{\ttfamily
  arXiv:1812.00211 [hep-ph]}}.

\bibitem{Hashiba:2018tbu}
S.~Hashiba and J.~Yokoyama, ``{Gravitational particle creation for dark matter
  and reheating},'' \href{http://dx.doi.org/10.1103/PhysRevD.99.043008}{{\em
  Phys. Rev. D} {\bfseries 99} no.~4, (2019) 043008},
  \href{http://arxiv.org/abs/1812.10032}{{\ttfamily arXiv:1812.10032
  [hep-ph]}}.

\bibitem{Ema:2019yrd}
Y.~Ema, K.~Nakayama, and Y.~Tang, ``{Production of purely gravitational dark
  matter: the case of fermion and vector boson},''
  \href{http://dx.doi.org/10.1007/JHEP07(2019)060}{{\em JHEP} {\bfseries 07}
  (2019) 060}, \href{http://arxiv.org/abs/1903.10973}{{\ttfamily
  arXiv:1903.10973 [hep-ph]}}.

\bibitem{Li:2019ves}
L.~Li, T.~Nakama, C.~M. Sou, Y.~Wang, and S.~Zhou, ``{Gravitational Production
  of Superheavy Dark Matter and Associated Cosmological Signatures},''
  \href{http://dx.doi.org/10.1007/JHEP07(2019)067}{{\em JHEP} {\bfseries 07}
  (2019) 067}, \href{http://arxiv.org/abs/1903.08842}{{\ttfamily
  arXiv:1903.08842 [astro-ph.CO]}}.

\bibitem{Cembranos:2019qlm}
J.~A.~R. Cembranos, L.~J. Garay, and J.~M. S\'anchez~Vel\'azquez,
  ``{Gravitational production of scalar dark matter},''
  \href{http://dx.doi.org/10.1007/JHEP06(2020)084}{{\em JHEP} {\bfseries 06}
  (2020) 084}, \href{http://arxiv.org/abs/1910.13937}{{\ttfamily
  arXiv:1910.13937 [hep-ph]}}.

\bibitem{Herring:2020cah}
N.~Herring and D.~Boyanovsky, ``{Gravitational production of nearly thermal
  fermionic dark matter},''
  \href{http://dx.doi.org/10.1103/PhysRevD.101.123522}{{\em Phys. Rev. D}
  {\bfseries 101} no.~12, (2020) 123522},
  \href{http://arxiv.org/abs/2005.00391}{{\ttfamily arXiv:2005.00391
  [astro-ph.CO]}}.

\bibitem{Ahmed:2020fhc}
A.~Ahmed, B.~Grzadkowski, and A.~Socha, ``{Gravitational production of vector
  dark matter},'' \href{http://dx.doi.org/10.1007/JHEP08(2020)059}{{\em JHEP}
  {\bfseries 08} (2020) 059}, \href{http://arxiv.org/abs/2005.01766}{{\ttfamily
  arXiv:2005.01766 [hep-ph]}}.

\bibitem{Ema:2020ggo}
Y.~Ema, R.~Jinno, and K.~Nakayama, ``{High-frequency Graviton from Inflaton
  Oscillation},'' \href{http://dx.doi.org/10.1088/1475-7516/2020/09/015}{{\em
  JCAP} {\bfseries 09} (2020) 015},
  \href{http://arxiv.org/abs/2006.09972}{{\ttfamily arXiv:2006.09972
  [astro-ph.CO]}}.

\bibitem{Karam:2020rpa}
A.~Karam, M.~Raidal, and E.~Tomberg, ``{Gravitational dark matter production in
  Palatini preheating},''
  \href{http://dx.doi.org/10.1088/1475-7516/2021/03/064}{{\em JCAP} {\bfseries
  03} (2021) 064}, \href{http://arxiv.org/abs/2007.03484}{{\ttfamily
  arXiv:2007.03484 [astro-ph.CO]}}.

\bibitem{Kolb:2020fwh}
E.~W. Kolb and A.~J. Long, ``{Completely dark photons from gravitational
  particle production during the inflationary era},''
  \href{http://dx.doi.org/10.1007/JHEP03(2021)283}{{\em JHEP} {\bfseries 03}
  (2021) 283}, \href{http://arxiv.org/abs/2009.03828}{{\ttfamily
  arXiv:2009.03828 [astro-ph.CO]}}.

\bibitem{Gross:2020zam}
C.~Gross, S.~Karamitsos, G.~Landini, and A.~Strumia, ``{Gravitational Vector
  Dark Matter},'' \href{http://dx.doi.org/10.1007/JHEP03(2021)174}{{\em JHEP}
  {\bfseries 03} (2021) 174}, \href{http://arxiv.org/abs/2012.12087}{{\ttfamily
  arXiv:2012.12087 [hep-ph]}}.

\bibitem{Ling:2021zlj}
S.~Ling and A.~J. Long, ``{Superheavy scalar dark matter from gravitational
  particle production in $\alpha$-attractor models of inflation},''
  \href{http://dx.doi.org/10.1103/PhysRevD.103.103532}{{\em Phys. Rev. D}
  {\bfseries 103} no.~10, (2021) 103532},
  \href{http://arxiv.org/abs/2101.11621}{{\ttfamily arXiv:2101.11621
  [astro-ph.CO]}}.

\bibitem{Mambrini:2021zpp}
Y.~Mambrini and K.~A. Olive, ``{Gravitational Production of Dark Matter during
  Reheating},'' \href{http://dx.doi.org/10.1103/PhysRevD.103.115009}{{\em Phys.
  Rev. D} {\bfseries 103} no.~11, (2021) 115009},
  \href{http://arxiv.org/abs/2102.06214}{{\ttfamily arXiv:2102.06214
  [hep-ph]}}.

\bibitem{Basso:2021whd}
E.~E. Basso and D.~J.~H. Chung, ``{Computation of gravitational particle
  production using adiabatic invariants},''
  \href{http://dx.doi.org/10.1007/JHEP11(2021)146}{{\em JHEP} {\bfseries 11}
  (2021) 146}, \href{http://arxiv.org/abs/2108.01653}{{\ttfamily
  arXiv:2108.01653 [hep-ph]}}.

\bibitem{Antoniadis:1993jc}
I.~Antoniadis, J.~Rizos, and K.~Tamvakis, ``{Singularity - free cosmological
  solutions of the superstring effective action},''
  \href{http://dx.doi.org/10.1016/0550-3213(94)90120-1}{{\em Nucl. Phys. B}
  {\bfseries 415} (1994) 497--514},
  \href{http://arxiv.org/abs/hep-th/9305025}{{\ttfamily arXiv:hep-th/9305025}}.

\bibitem{Kanti:1998jd}
P.~Kanti, J.~Rizos, and K.~Tamvakis, ``{Singularity free cosmological solutions
  in quadratic gravity},''
  \href{http://dx.doi.org/10.1103/PhysRevD.59.083512}{{\em Phys. Rev. D}
  {\bfseries 59} (1999) 083512},
  \href{http://arxiv.org/abs/gr-qc/9806085}{{\ttfamily arXiv:gr-qc/9806085}}.

\bibitem{Nojiri:2005vv}
S.~Nojiri, S.~D. Odintsov, and M.~Sasaki, ``{Gauss-Bonnet dark energy},''
  \href{http://dx.doi.org/10.1103/PhysRevD.71.123509}{{\em Phys. Rev. D}
  {\bfseries 71} (2005) 123509},
  \href{http://arxiv.org/abs/hep-th/0504052}{{\ttfamily arXiv:hep-th/0504052}}.

\bibitem{Lue:1998mq}
A.~Lue, L.-M. Wang, and M.~Kamionkowski, ``{Cosmological signature of new
  parity violating interactions},''
  \href{http://dx.doi.org/10.1103/PhysRevLett.83.1506}{{\em Phys. Rev. Lett.}
  {\bfseries 83} (1999) 1506--1509},
  \href{http://arxiv.org/abs/astro-ph/9812088}{{\ttfamily
  arXiv:astro-ph/9812088}}.

\bibitem{Choi:1999zy}
K.~Choi, J.-c. Hwang, and K.~W. Hwang, ``{String theoretic axion coupling and
  the evolution of cosmic structures},''
  \href{http://dx.doi.org/10.1103/PhysRevD.61.084026}{{\em Phys. Rev. D}
  {\bfseries 61} (2000) 084026},
  \href{http://arxiv.org/abs/hep-ph/9907244}{{\ttfamily arXiv:hep-ph/9907244}}.

\bibitem{Alexander:2004us}
S.~H.-S. Alexander, M.~E. Peskin, and M.~M. Sheikh-Jabbari, ``{Leptogenesis
  from gravity waves in models of inflation},''
  \href{http://dx.doi.org/10.1103/PhysRevLett.96.081301}{{\em Phys. Rev. Lett.}
  {\bfseries 96} (2006) 081301},
  \href{http://arxiv.org/abs/hep-th/0403069}{{\ttfamily arXiv:hep-th/0403069}}.

\bibitem{Alexander:2004wk}
S.~Alexander and J.~Martin, ``{Birefringent gravitational waves and the
  consistency check of inflation},''
  \href{http://dx.doi.org/10.1103/PhysRevD.71.063526}{{\em Phys. Rev. D}
  {\bfseries 71} (2005) 063526},
  \href{http://arxiv.org/abs/hep-th/0410230}{{\ttfamily arXiv:hep-th/0410230}}.

\bibitem{Lyth:2005jf}
D.~H. Lyth, C.~Quimbay, and Y.~Rodriguez, ``{Leptogenesis and tensor
  polarisation from a gravitational Chern-Simons term},''
  \href{http://dx.doi.org/10.1088/1126-6708/2005/03/016}{{\em JHEP} {\bfseries
  03} (2005) 016}, \href{http://arxiv.org/abs/hep-th/0501153}{{\ttfamily
  arXiv:hep-th/0501153}}.

\bibitem{Fischler:2007tj}
W.~Fischler and S.~Paban, ``{Leptogenesis from Pseudo-Scalar Driven
  Inflation},'' \href{http://dx.doi.org/10.1088/1126-6708/2007/10/066}{{\em
  JHEP} {\bfseries 10} (2007) 066},
  \href{http://arxiv.org/abs/0708.3828}{{\ttfamily arXiv:0708.3828 [hep-th]}}.

\bibitem{DeSimone:2016bok}
A.~De~Simone, T.~Kobayashi, and S.~Liberati, ``{Geometric Baryogenesis from
  Shift Symmetry},''
  \href{http://dx.doi.org/10.1103/PhysRevLett.118.131101}{{\em Phys. Rev.
  Lett.} {\bfseries 118} no.~13, (2017) 131101},
  \href{http://arxiv.org/abs/1612.04824}{{\ttfamily arXiv:1612.04824
  [hep-ph]}}.

\bibitem{Kawai:2017kqt}
S.~Kawai and J.~Kim, ``{Gauss\textendash{}Bonnet Chern\textendash{}Simons
  gravitational wave leptogenesis},''
  \href{http://dx.doi.org/10.1016/j.physletb.2018.12.019}{{\em Phys. Lett. B}
  {\bfseries 789} (2019) 145--149},
  \href{http://arxiv.org/abs/1702.07689}{{\ttfamily arXiv:1702.07689
  [hep-th]}}.

\bibitem{Kamada:2020jaf}
K.~Kamada, J.~Kume, and Y.~Yamada, ``{Renormalization in gravitational
  leptogenesis with pseudo-scalar-tensor coupling},''
  \href{http://dx.doi.org/10.1088/1475-7516/2020/10/030}{{\em JCAP} {\bfseries
  10} (2020) 030}, \href{http://arxiv.org/abs/2007.08029}{{\ttfamily
  arXiv:2007.08029 [hep-ph]}}.

\bibitem{Delbourgo:2000nq}
R.~Delbourgo and D.-s. Liu, ``{Electromagnetic and gravitational decay of the
  Higgs boson},'' {\em Austral. J. Phys.} {\bfseries 53} (2001) 647--651,
  \href{http://arxiv.org/abs/hep-ph/0004156}{{\ttfamily arXiv:hep-ph/0004156}}.

\bibitem{Alonzo-Artiles:2021mym}
A.~Alonzo-Artiles, A.~Avilez-L\'opez, J.~L. D\'\i{}az-Cruz, and B.~O.
  Larios-L\'opez, ``{The Higgs-Graviton Couplings: from Amplitudes to the
  Action},'' \href{http://arxiv.org/abs/2105.11684}{{\ttfamily arXiv:2105.11684
  [hep-th]}}.

\bibitem{LIGOScientific:2016aoc}
{\bfseries LIGO Scientific, Virgo} Collaboration, B.~P. Abbott {\em et~al.},
  ``{Observation of Gravitational Waves from a Binary Black Hole Merger},''
  \href{http://dx.doi.org/10.1103/PhysRevLett.116.061102}{{\em Phys. Rev.
  Lett.} {\bfseries 116} no.~6, (2016) 061102},
  \href{http://arxiv.org/abs/1602.03837}{{\ttfamily arXiv:1602.03837 [gr-qc]}}.

\bibitem{Ghiglieri:2015nfa}
J.~Ghiglieri and M.~Laine, ``{Gravitational wave background from Standard Model
  physics: Qualitative features},''
  \href{http://dx.doi.org/10.1088/1475-7516/2015/07/022}{{\em JCAP} {\bfseries
  07} (2015) 022}, \href{http://arxiv.org/abs/1504.02569}{{\ttfamily
  arXiv:1504.02569 [hep-ph]}}.

\bibitem{Ghiglieri:2020mhm}
J.~Ghiglieri, G.~Jackson, M.~Laine, and Y.~Zhu, ``{Gravitational wave
  background from Standard Model physics: Complete leading order},''
  \href{http://dx.doi.org/10.1007/JHEP07(2020)092}{{\em JHEP} {\bfseries 07}
  (2020) 092}, \href{http://arxiv.org/abs/2004.11392}{{\ttfamily
  arXiv:2004.11392 [hep-ph]}}.

\bibitem{Ringwald:2020ist}
A.~Ringwald, J.~Sch\"utte-Engel, and C.~Tamarit, ``{Gravitational Waves as a
  Big Bang Thermometer},''
  \href{http://dx.doi.org/10.1088/1475-7516/2021/03/054}{{\em JCAP} {\bfseries
  03} (2021) 054}, \href{http://arxiv.org/abs/2011.04731}{{\ttfamily
  arXiv:2011.04731 [hep-ph]}}.

\bibitem{Nakayama:2018ptw}
K.~Nakayama and Y.~Tang, ``{Stochastic Gravitational Waves from Particle
  Origin},'' \href{http://dx.doi.org/10.1016/j.physletb.2018.11.023}{{\em Phys.
  Lett. B} {\bfseries 788} (2019) 341--346},
  \href{http://arxiv.org/abs/1810.04975}{{\ttfamily arXiv:1810.04975
  [hep-ph]}}.

\bibitem{Huang:2019lgd}
D.~Huang and L.~Yin, ``{Stochastic Gravitational Waves from Inflaton Decays},''
  \href{http://dx.doi.org/10.1103/PhysRevD.100.043538}{{\em Phys. Rev. D}
  {\bfseries 100} no.~4, (2019) 043538},
  \href{http://arxiv.org/abs/1905.08510}{{\ttfamily arXiv:1905.08510
  [hep-ph]}}.

\bibitem{Maggiore:1999vm}
M.~Maggiore, ``{Gravitational wave experiments and early universe cosmology},''
  \href{http://dx.doi.org/10.1016/S0370-1573(99)00102-7}{{\em Phys. Rept.}
  {\bfseries 331} (2000) 283--367},
  \href{http://arxiv.org/abs/gr-qc/9909001}{{\ttfamily arXiv:gr-qc/9909001}}.

\bibitem{Sabin:2014bua}
C.~Sabin, D.~E. Bruschi, M.~Ahmadi, and I.~Fuentes, ``{Phonon creation by
  gravitational waves},''
  \href{http://dx.doi.org/10.1088/1367-2630/16/8/085003}{{\em New J. Phys.}
  {\bfseries 16} (2014) 085003},
  \href{http://arxiv.org/abs/1402.7009}{{\ttfamily arXiv:1402.7009
  [quant-ph]}}.

\bibitem{Robbins:2021ucj}
M.~P.~G. Robbins, N.~Afshordi, A.~O. Jamison, and R.~B. Mann, ``{Detection of
  Gravitational Waves using Parametric Resonance in Bose-Einstein
  Condensates},'' \href{http://arxiv.org/abs/2101.03691}{{\ttfamily
  arXiv:2101.03691 [gr-qc]}}.

\bibitem{Ito:2019wcb}
A.~Ito, T.~Ikeda, K.~Miuchi, and J.~Soda, ``{Probing GHz gravitational waves
  with graviton\textendash{}magnon resonance},''
  \href{http://dx.doi.org/10.1140/epjc/s10052-020-7735-y}{{\em Eur. Phys. J. C}
  {\bfseries 80} no.~3, (2020) 179},
  \href{http://arxiv.org/abs/1903.04843}{{\ttfamily arXiv:1903.04843 [gr-qc]}}.

\bibitem{Ito:2020wxi}
A.~Ito and J.~Soda, ``{A formalism for magnon gravitational wave detectors},''
  \href{http://dx.doi.org/10.1140/epjc/s10052-020-8092-6}{{\em Eur. Phys. J. C}
  {\bfseries 80} no.~6, (2020) 545},
  \href{http://arxiv.org/abs/2004.04646}{{\ttfamily arXiv:2004.04646 [gr-qc]}}.

\bibitem{Aggarwal:2020olq}
N.~Aggarwal {\em et~al.}, ``{Challenges and Opportunities of Gravitational Wave
  Searches at MHz to GHz Frequencies},''
  \href{http://arxiv.org/abs/2011.12414}{{\ttfamily arXiv:2011.12414 [gr-qc]}}.

\bibitem{Berlin:2021txa}
A.~Berlin, D.~Blas, R.~T. D'Agnolo, S.~A.~R. Ellis, R.~Harnik, Y.~Kahn, and
  J.~Sch\"utte-Engel, ``{Detecting High-Frequency Gravitational Waves with
  Microwave Cavities},'' \href{http://arxiv.org/abs/2112.11465}{{\ttfamily
  arXiv:2112.11465 [hep-ph]}}.

\bibitem{DeWitt:1975ys}
B.~S. DeWitt, ``{Quantum Field Theory in Curved Space-Time},''
  \href{http://dx.doi.org/10.1016/0370-1573(75)90051-4}{{\em Phys. Rept.}
  {\bfseries 19} (1975) 295--357}.

\bibitem{Birrell:1982ix}
N.~D. Birrell and P.~C.~W. Davies,
  \href{http://dx.doi.org/10.1017/CBO9780511622632}{{\em {Quantum Fields in
  Curved Space}}}.
\newblock Cambridge Monographs on Mathematical Physics. Cambridge Univ. Press,
  Cambridge, UK, 2, 1984.

\bibitem{Parker:2009uva}
L.~E. Parker and D.~Toms,
  \href{http://dx.doi.org/10.1017/CBO9780511813924}{{\em {Quantum Field Theory
  in Curved Spacetime}: {Quantized Field and Gravity}}}.
\newblock Cambridge Monographs on Mathematical Physics. Cambridge University
  Press, 8, 2009.

\bibitem{Parker:1984dj}
L.~Parker and D.~J. Toms, ``{New Form for the Coincidence Limit of the Feynman
  Propagator, or Heat Kernel, in Curved Space-time},''
  \href{http://dx.doi.org/10.1103/PhysRevD.31.953}{{\em Phys. Rev. D}
  {\bfseries 31} (1985) 953}.

\bibitem{Jack:1985mw}
I.~Jack and L.~Parker, ``{Proof of Summed Form of Proper Time Expansion for
  Propagator in Curved Space-time},''
  \href{http://dx.doi.org/10.1103/PhysRevD.31.2439}{{\em Phys. Rev. D}
  {\bfseries 31} (1985) 2439}.

\bibitem{Christensen:1976vb}
S.~M. Christensen, ``{Vacuum Expectation Value of the Stress Tensor in an
  Arbitrary Curved Background: The Covariant Point Separation Method},''
  \href{http://dx.doi.org/10.1103/PhysRevD.14.2490}{{\em Phys. Rev. D}
  {\bfseries 14} (1976) 2490--2501}.

\bibitem{Avramidi:1990je}
I.~G. Avramidi, ``{The Covariant Technique for Calculation of One Loop
  Effective Action},''
  \href{http://dx.doi.org/10.1016/0550-3213(91)90492-G}{{\em Nucl. Phys. B}
  {\bfseries 355} (1991) 712--754}. [Erratum: Nucl.Phys.B 509, 557--558
  (1998)].

\bibitem{Fujikawa:1980eg}
K.~Fujikawa, ``{Path Integral for Gauge Theories with Fermions},''
  \href{http://dx.doi.org/10.1103/PhysRevD.21.2848}{{\em Phys. Rev. D}
  {\bfseries 21} (1980) 2848}. [Erratum: Phys.Rev.D 22, 1499 (1980)].

\bibitem{Georgi:1986df}
H.~Georgi, D.~B. Kaplan, and L.~Randall, ``{Manifesting the Invisible Axion at
  Low-energies},'' \href{http://dx.doi.org/10.1016/0370-2693(86)90688-X}{{\em
  Phys. Lett. B} {\bfseries 169} (1986) 73--78}.

\bibitem{Shifman:1979eb}
M.~A. Shifman, A.~I. Vainshtein, M.~B. Voloshin, and V.~I. Zakharov,
  ``{Low-Energy Theorems for Higgs Boson Couplings to Photons},'' {\em Sov. J.
  Nucl. Phys.} {\bfseries 30} (1979) 711--716.

\bibitem{Salvio:2018crh}
A.~Salvio, ``{Quadratic Gravity},''
  \href{http://dx.doi.org/10.3389/fphy.2018.00077}{{\em Front. in Phys.}
  {\bfseries 6} (2018) 77}, \href{http://arxiv.org/abs/1804.09944}{{\ttfamily
  arXiv:1804.09944 [hep-th]}}.

\bibitem{Salvio:2014soa}
A.~Salvio and A.~Strumia, ``{Agravity},''
  \href{http://dx.doi.org/10.1007/JHEP06(2014)080}{{\em JHEP} {\bfseries 06}
  (2014) 080}, \href{http://arxiv.org/abs/1403.4226}{{\ttfamily arXiv:1403.4226
  [hep-ph]}}.

\bibitem{Salvio:2017qkx}
A.~Salvio and A.~Strumia, ``{Agravity up to infinite energy},''
  \href{http://dx.doi.org/10.1140/epjc/s10052-018-5588-4}{{\em Eur. Phys. J. C}
  {\bfseries 78} no.~2, (2018) 124},
  \href{http://arxiv.org/abs/1705.03896}{{\ttfamily arXiv:1705.03896
  [hep-th]}}.

\bibitem{Fradkin:1981iu}
E.~S. Fradkin and A.~A. Tseytlin, ``{Renormalizable asymptotically free quantum
  theory of gravity},''
  \href{http://dx.doi.org/10.1016/0550-3213(82)90444-8}{{\em Nucl. Phys. B}
  {\bfseries 201} (1982) 469--491}.

\bibitem{Buchbinder:1992rb}
I.~L. Buchbinder, S.~D. Odintsov, and I.~L. Shapiro, {\em {Effective action in
  quantum gravity}}.
\newblock 1992.

\bibitem{Elizalde:1993ee}
E.~Elizalde and S.~D. Odintsov, ``{Renormalization group improved effective
  potential for gauge theories in curved space-time},''
  \href{http://dx.doi.org/10.1016/0370-2693(93)91427-O}{{\em Phys. Lett. B}
  {\bfseries 303} (1993) 240--248},
  \href{http://arxiv.org/abs/hep-th/9302074}{{\ttfamily arXiv:hep-th/9302074}}.

\bibitem{Elizalde:1993ew}
E.~Elizalde and S.~D. Odintsov, ``{Renormalization group improved effective
  Lagrangian for interacting theories in curved space-time},''
  \href{http://dx.doi.org/10.1016/0370-2693(94)90464-2}{{\em Phys. Lett. B}
  {\bfseries 321} (1994) 199--204},
  \href{http://arxiv.org/abs/hep-th/9311087}{{\ttfamily arXiv:hep-th/9311087}}.

\bibitem{Codello:2015mba}
A.~Codello and R.~K. Jain, ``{On the covariant formalism of the effective field
  theory of gravity and leading order corrections},''
  \href{http://dx.doi.org/10.1088/0264-9381/33/22/225006}{{\em Class. Quant.
  Grav.} {\bfseries 33} no.~22, (2016) 225006},
  \href{http://arxiv.org/abs/1507.06308}{{\ttfamily arXiv:1507.06308 [gr-qc]}}.

\bibitem{Markkanen:2018bfx}
T.~Markkanen, S.~Nurmi, A.~Rajantie, and S.~Stopyra, ``{The 1-loop effective
  potential for the Standard Model in curved spacetime},''
  \href{http://dx.doi.org/10.1007/JHEP06(2018)040}{{\em JHEP} {\bfseries 06}
  (2018) 040}, \href{http://arxiv.org/abs/1804.02020}{{\ttfamily
  arXiv:1804.02020 [hep-ph]}}.

\bibitem{Zee:1978wi}
A.~Zee, ``{A Broken Symmetric Theory of Gravity},''
  \href{http://dx.doi.org/10.1103/PhysRevLett.42.417}{{\em Phys. Rev. Lett.}
  {\bfseries 42} (1979) 417}.

\bibitem{Barr:1989jn}
S.~M. Barr and G.~Segre, ``{Limits on Reheating in Inflationary Cosmology
  Models},'' \href{http://dx.doi.org/10.1103/PhysRevLett.62.2781}{{\em Phys.
  Rev. Lett.} {\bfseries 62} (1989) 2781--2784}.

\bibitem{Cervantes-Cota:1995ehs}
J.~L. Cervantes-Cota and H.~Dehnen, ``{Induced gravity inflation in the
  standard model of particle physics},''
  \href{http://dx.doi.org/10.1016/0550-3213(95)00128-X}{{\em Nucl. Phys. B}
  {\bfseries 442} (1995) 391--412},
  \href{http://arxiv.org/abs/astro-ph/9505069}{{\ttfamily
  arXiv:astro-ph/9505069}}.

\bibitem{Saikawa:2018rcs}
K.~Saikawa and S.~Shirai, ``{Primordial gravitational waves, precisely: The
  role of thermodynamics in the Standard Model},''
  \href{http://dx.doi.org/10.1088/1475-7516/2018/05/035}{{\em JCAP} {\bfseries
  05} (2018) 035}, \href{http://arxiv.org/abs/1803.01038}{{\ttfamily
  arXiv:1803.01038 [hep-ph]}}.

\bibitem{Kallosh:2013hoa}
R.~Kallosh and A.~Linde, ``{Universality Class in Conformal Inflation},''
  \href{http://dx.doi.org/10.1088/1475-7516/2013/07/002}{{\em JCAP} {\bfseries
  07} (2013) 002}, \href{http://arxiv.org/abs/1306.5220}{{\ttfamily
  arXiv:1306.5220 [hep-th]}}.

\bibitem{Ferrara:2013rsa}
S.~Ferrara, R.~Kallosh, A.~Linde, and M.~Porrati, ``{Minimal Supergravity
  Models of Inflation},''
  \href{http://dx.doi.org/10.1103/PhysRevD.88.085038}{{\em Phys. Rev. D}
  {\bfseries 88} no.~8, (2013) 085038},
  \href{http://arxiv.org/abs/1307.7696}{{\ttfamily arXiv:1307.7696 [hep-th]}}.

\bibitem{Kallosh:2013yoa}
R.~Kallosh, A.~Linde, and D.~Roest, ``{Superconformal Inflationary
  $\alpha$-Attractors},'' \href{http://dx.doi.org/10.1007/JHEP11(2013)198}{{\em
  JHEP} {\bfseries 11} (2013) 198},
  \href{http://arxiv.org/abs/1311.0472}{{\ttfamily arXiv:1311.0472 [hep-th]}}.

\bibitem{Galante:2014ifa}
M.~Galante, R.~Kallosh, A.~Linde, and D.~Roest, ``{Unity of Cosmological
  Inflation Attractors},''
  \href{http://dx.doi.org/10.1103/PhysRevLett.114.141302}{{\em Phys. Rev.
  Lett.} {\bfseries 114} no.~14, (2015) 141302},
  \href{http://arxiv.org/abs/1412.3797}{{\ttfamily arXiv:1412.3797 [hep-th]}}.

\bibitem{Ema:2016ops}
Y.~Ema, K.~Hamaguchi, T.~Moroi, and K.~Nakayama, ``{Flaxion: a minimal
  extension to solve puzzles in the standard model},''
  \href{http://dx.doi.org/10.1007/JHEP01(2017)096}{{\em JHEP} {\bfseries 01}
  (2017) 096}, \href{http://arxiv.org/abs/1612.05492}{{\ttfamily
  arXiv:1612.05492 [hep-ph]}}.

\bibitem{Ford:1986sy}
L.~H. Ford, ``{Gravitational Particle Creation and Inflation},''
  \href{http://dx.doi.org/10.1103/PhysRevD.35.2955}{{\em Phys. Rev. D}
  {\bfseries 35} (1987) 2955}.

\bibitem{Chung:1998zb}
D.~J.~H. Chung, E.~W. Kolb, and A.~Riotto, ``{Superheavy dark matter},''
  \href{http://dx.doi.org/10.1103/PhysRevD.59.023501}{{\em Phys. Rev. D}
  {\bfseries 59} (1998) 023501},
  \href{http://arxiv.org/abs/hep-ph/9802238}{{\ttfamily arXiv:hep-ph/9802238}}.

\bibitem{Adshead:2018oaa}
P.~Adshead, L.~Pearce, M.~Peloso, M.~A. Roberts, and L.~Sorbo, ``{Phenomenology
  of fermion production during axion inflation},''
  \href{http://dx.doi.org/10.1088/1475-7516/2018/06/020}{{\em JCAP} {\bfseries
  06} (2018) 020}, \href{http://arxiv.org/abs/1803.04501}{{\ttfamily
  arXiv:1803.04501 [astro-ph.CO]}}.

\bibitem{Ryu2012}
S.~Ryu, J.~E. Moore, and A.~W.~W. Ludwig, ``{Electromagnetic and gravitational
  responses and anomalies in topological insulators and superconductors},''
  \href{http://dx.doi.org/10.1103/PhysRevB.85.045104}{{\em Phys. Rev. B}
  {\bfseries 85} (2012) 045104},
  \href{http://arxiv.org/abs/1010.0936}{{\ttfamily arXiv:1010.0936
  [cond-mat.str-el]}}.

\bibitem{Wang:2010xh}
Z.~Wang, X.-L. Qi, and S.-C. Zhang, ``{Topological Field Theory and Thermal
  Responses of Interacting Topological Superconductors},''
  \href{http://dx.doi.org/10.1103/PhysRevB.84.014527}{{\em Phys. Rev. B}
  {\bfseries 84} (2011) 014527},
  \href{http://arxiv.org/abs/1011.0586}{{\ttfamily arXiv:1011.0586
  [cond-mat.str-el]}}.

\bibitem{Shiozaki:2013wda}
K.~Shiozaki and S.~Fujimoto, ``{Dynamical axion in topological superconductors
  and superfluids},'' \href{http://dx.doi.org/10.1103/PhysRevB.89.054506}{{\em
  Phys. Rev. B} {\bfseries 89} no.~5, (2014) 054506},
  \href{http://arxiv.org/abs/1310.4982}{{\ttfamily arXiv:1310.4982
  [cond-mat.supr-con]}}.

\bibitem{Banyeres:2018aax}
M.~Banyeres, G.~Dom\`enech, and J.~Garriga, ``{Vacuum birefringence and the
  Schwinger effect in (3+1) de Sitter},''
  \href{http://dx.doi.org/10.1088/1475-7516/2018/10/023}{{\em JCAP} {\bfseries
  10} (2018) 023}, \href{http://arxiv.org/abs/1809.08977}{{\ttfamily
  arXiv:1809.08977 [hep-th]}}.

\bibitem{Domcke:2019qmm}
V.~Domcke, Y.~Ema, and K.~Mukaida, ``{Chiral Anomaly, Schwinger Effect,
  Euler-Heisenberg Lagrangian, and application to axion inflation},''
  \href{http://dx.doi.org/10.1007/JHEP02(2020)055}{{\em JHEP} {\bfseries 02}
  (2020) 055}, \href{http://arxiv.org/abs/1910.01205}{{\ttfamily
  arXiv:1910.01205 [hep-ph]}}.

\bibitem{Ellis:2016jkw}
J.~Ellis, ``{TikZ-Feynman: Feynman diagrams with TikZ},''
  \href{http://dx.doi.org/10.1016/j.cpc.2016.08.019}{{\em Comput. Phys.
  Commun.} {\bfseries 210} (2017) 103--123},
  \href{http://arxiv.org/abs/1601.05437}{{\ttfamily arXiv:1601.05437
  [hep-ph]}}.

\bibitem{Poisson:2011nh}
E.~Poisson, A.~Pound, and I.~Vega, ``{The Motion of point particles in curved
  spacetime},'' \href{http://dx.doi.org/10.12942/lrr-2011-7}{{\em Living Rev.
  Rel.} {\bfseries 14} (2011) 7},
  \href{http://arxiv.org/abs/1102.0529}{{\ttfamily arXiv:1102.0529 [gr-qc]}}.

\end{thebibliography}\endgroup

\end{document}